\documentclass[12pt, openright, oneside, a4paper, english]{abntex2}

\usepackage{lmodern}			
\usepackage[T1]{fontenc}		
\usepackage[utf8]{inputenc}		
\usepackage{lastpage}			
\usepackage{indentfirst}		
\usepackage{color}				
\usepackage{graphicx}			
\usepackage{microtype} 			
\usepackage{amsthm}
\usepackage{amsmath}
\usepackage{mathtools}
\usepackage{amsfonts}
\usepackage{amssymb}
\usepackage{caption}
\usepackage{subcaption}
\usepackage{tabularx}
\usepackage{csquotes}
\usepackage{makecell}
\usepackage{siunitx}

\usepackage{amsmath}

\usepackage[ruled,lined,linesnumbered]{algorithm2e}

\usepackage{multirow}

\usepackage[brazil]{babel}
\addto\captionsbrazil{%

\renewcommand{\listfigurename}{List of Figures}

\renewcommand{\contentsname}{Contents}

}

\usepackage{lipsum}				

\titulo{Anomalous Transport In Low Dimension Materials}
\autor{Leonardo Arruda Lopes}
\local{São Paulo}
\data{2025}
\orientador{Prof. Dr. Ana Júlia Silveira Mizher}

\instituicao{%
 University of Cidade de São Paulo
  \par
 Master and Doctoral program in Astrophysics and Computational Physics}
\tipotrabalho{Master Thesis}
\preambulo{Master Thesis presented to the Master and Doctoral program in Astrophysics and Computational Physics of the University of Cidade de São Paulo as a partial requirement to obtain the degree of Master in Theoretical Physics}

\definecolor{blue}{RGB}{41,5,195}

\makeatletter
\hypersetup{
		pdftitle={\@title}, 
		pdfauthor={\@author},
    	pdfsubject={\imprimirpreambulo},
	    pdfcreator={LaTeX with abnTeX2},
		pdfkeywords={abnt}{latex}{abntex}{abntex2}{trabalho acadêmico}, 
		colorlinks=true,       		
    	linkcolor=blue,          	
    	citecolor=blue,        		
    	filecolor=magenta,      		
		urlcolor=blue,
		bookmarksdepth=4
}
\makeatother

\makeindex

\begin{document}

\frenchspacing 

\pretextual

\imprimircapa
\cleardoublepage

\imprimirfolhaderosto



\newpage
\begin{center}
{\ABNTEXchapterfont\Large\textsc{Statement of Authorship}}
\end{center}

\vspace*{1cm}

I hereby declare that the thesis submitted is my own work. All direct or indirect sources used are acknowledged as references. I further declare that I have not submitted this thesis at any other institution in order to obtain a degree. \\


\newpage
\begin{dedicatoria}
  \vspace*{\fill}
  \centering
  \noindent
  \textit{To my beloved wife, who supported and encouraged me throughout the process.\\ To my dear advisor, who, with great patience and a desire to help, got me to where I am today.\\ This work is yours too.} \vspace*{\fill}
\end{dedicatoria}




	
		

\newpage
\setlength{\absparsep}{18pt} 
\begin{resumo}[Resumo]

Esta dissertação apresenta uma investigação teórica sistemática sobre a realização de um análogo de matéria condensada do Efeito Quiral Magnético (EQM) num sistema quasi-planar $2+1D$. A pesquisa estabelece uma ponte conceitual entre os fenômenos de transporte anômalo da física de altas energias e as propriedades eletrônicas emergentes de redes de favo de mel projetadas. O objetivo central é a formulação de um Hamiltoniano efetivo de baixa energia que incorpore os ingredientes necessários para um efeito do tipo EQM. Isso é alcançado indo além do grafeno puro, cuja simetria de sub-rede inerente impede a formação de um gap de massa necessário para definir estados pseudo-quirais robustos. O cerne deste trabalho é um modelo baseado numa rede de favo de mel com simetria de sub-rede explicitamente quebrada, que introduz um gap de banda e dota as quasipartículas do sistema com uma pseudo-quiralidade bem definida baseada na polarização da sub-rede. Um parâmetro de quebra de simetria de reversão temporal é introduzido para modificar assimetricamente os gaps dos vales, criando um desequilíbrio de não-equilíbrio controlável análogo ao potencial químico quiral em sistemas relativísticos. Um achado fundamental é a validação da consistência física do modelo; através de cálculos de comutadores, o momento angular total — compreendendo tanto a componente orbital quanto uma componente emergente de ``spin de rede'' — mostra-se uma quantidade conservada. Esta pesquisa transforma com sucesso a possibilidade abstrata de um EQM 2D num quadro teórico concreto e autoconsistente, detalhando as condições de simetria precisas necessárias para a sua manifestação.

\textbf{Palavras-chave:} Efeito Quiral Magnético, Quebra de Simetria de Sub-rede, Rede hexagonal, Transporte Anômalo, Spin de Rede

\end{resumo}

\newpage

\setlength{\absparsep}{18pt} 
\begin{resumo}[Abstract]
 \begin{otherlanguage*}{english}

This dissertation presents a systematic theoretical investigation into realizing a condensed matter analogue of the Chiral Magnetic Effect (CME) in a quasi-planar, $2+1D$ system. The research establishes a conceptual bridge between the anomalous transport phenomena of high-energy physics and the emergent electronic properties of engineered honeycomb lattices. The central objective is the formulation of a low-energy effective Hamiltonian that incorporates the necessary ingredients for a CME - like effect. This is achieved by moving beyond pristine graphene, whose inherent sublattice symmetry precludes the formation of a mass gap necessary for defining robust pseudo-chiral states. The core of this work is a model based on a honeycomb lattice with explicitly broken sublattice symmetry, which introduces a band gap and endows the quasi-particles system with a well-defined pseudo-chirality based on sublattice polarization. A time-reversal symmetry-breaking parameter is introduced to asymmetrically modify the valley gaps, creating a controllable non-equilibrium imbalance analogous to the chiral chemical potential in relativistic systems. A key finding is the validation of the physical model consistency; through commutator calculations, the total angular momentum — comprising both orbital and an emergent ``lattice spin'' component — is shown to be a conserved quantity. This research successfully transforms the abstract possibility of a 2D CME into a concrete, self-consistent theoretical framework, detailing the precise symmetry conditions required for its manifestation.


\end{otherlanguage*}
\end{resumo}

\newpage

\pdfbookmark[0]{\listfigurename}{lof}
\listoffigures*
\cleardoublepage


\pdfbookmark[0]{\contentsname}{toc}
\tableofcontents*
\cleardoublepage

\textual

\chapter{Introduction}
The disparate domains of high-energy particle physics and condensed matter physics have traditionally pursued independent research trajectories, each defined by vastly different energy scales and physical systems. The former investigates the fundamental constituents of matter under extreme conditions, such as the deconfined state of quarks and gluons in the Quark-Gluon Plasma (QGP), while the latter explores the emergent collective behaviors of electrons within the highly ordered environment of crystalline solids. Despite these apparent differences, a central thesis of modern theoretical physics is that the underlying principles of quantum field theory provide a universal descriptive framework. This universality suggests that phenomena once thought to be exclusive to one domain may have direct analogues in the other \cite{Katsnelson_2006,VOZMEDIANO2010109,Cortijo_2012,Chernodub_2014,Iorio_2012,Iorio_2014,Li2016,Dudal:2021ret}. This dissertation is dedicated to the construction and exploration of one such theoretical bridge. Specifically, it investigates the possibility of realizing a phenomenon known as the Chiral Magnetic Effect (CME) — an anomalous transport effect rooted in the topological structure of Quantum Chromodynamics (QCD) \cite{KHARZEEV2008227,Fukushima:2008xe} — within the context of quasi-planar, two-dimensional condensed matter systems. By systematically developing a model for engineered honeycomb lattices, this work aims to demonstrate that the fundamental symmetries and quantum anomalies that give rise to the CME in the relativistic environment of the QGP can be replicated and observed through the behavior of quasiparticle excitations in a solid-state material .

The CME is a macroscopic quantum phenomenon that originates from the complex vacuum structure of QCD, the theory of the strong interaction. The QCD vacuum is not an empty void but a dynamic medium, characterized by both non-topological fluctuations of quark-antiquark pairs and gluon fields that form instanton-like configurations with a non-trivial topological structure \cite{KHARZEEV2008227,Cheng:1984vwu}. Under the extreme conditions of temperature and density found in the QGP — a state of matter thought to have existed in the first microseconds after the Big Bang and recreated today in particle accelerators — these topological fluctuations can lead to a localized imbalance in the number of left-handed and right-handed quarks. This property, known as chirality or ``handedness'', is a fundamental aspect of relativistic fermions. In the presence of the immensely powerful magnetic fields generated during off-center heavy-ion collisions \cite{Adhikari:2024bfa}, this temporary chiral imbalance is predicted to induce a separation of electric charge, generating an electric current that flows parallel to the magnetic field direction. The observation of the CME in the QGP would provide direct evidence of the topological nature of the strong force and local violations of fundamental symmetries, offering a unique window into the early universe.

For years, this captivating effect was considered exclusive to the domain of high-energy physics. However, a revolutionary insight bridged this gap: the fundamental ingredients required for the CME — massless, chiral fermions and a mechanism for inducing a chiral imbalance — are not unique to the QGP. They find a natural home in a class of recently discovered topological materials known as Dirac and Weyl semimetals \cite{Li2016,PhysRevB.93.115414,behnami2025chiralanomalyweylsemimetal}. These three-dimensional materials possess a unique electronic band structure where the conduction and valence bands touch at discrete points, called Dirac or Weyl nodes. Near these nodes, the electronic material excitations, or quasi-particles, behave not as conventional electrons but as relativistic massless Dirac or Weyl fermions. This emergent relativistic behavior means that the complex physics of quantum field theory can be simulated and studied on a laboratory tabletop.

In these materials, the Weyl nodes come in pairs of opposite chirality. The application of parallel electric and magnetic fields can ``pump'' charge between nodes of opposite chirality, creating the necessary chiral imbalance. This, in turn, generates the anomalous CME current, which manifests experimentally as a large, negative longitudinal magnetoresistance — a decrease in electrical resistance when the magnetic field is aligned with the current. The stunning experimental confirmation of this effect in materials like $ZrTe_{5}$ \cite{Li2016} and $TaAs$ \cite{PhysRevX.5.031013} marked a landmark achievement, proving that the physics of the chiral anomaly could be harnessed in a solid-state system.

The ``successful observation'' of the CME in $3+1 D $ systems, predicted to occur in the QGP and observed in Weyl semimetals \cite{Li2016,PhysRevX.5.031013}, naturally raises the next fundamental question, which forms the central inquiry of this dissertation: can this intrinsically three-dimensional effect manifest in systems of lower dimensionality? The quintessential $2+1D$ material, graphene, immediately presents itself as the primary candidate. Its celebrated electronic structure, featuring two inequivalent ``valleys'' ($K^{+}$ and $K^{-}$) where electrons behave as massless Dirac fermions, suggests a direct parallel with the chiral fermions required for the CME, with the valley index playing a role analogous to chirality.

However, pristine graphene presents a fundamental obstacle rooted in its high degree of symmetry. The perfect inversion symmetry between its two carbon sublattices ($A$ and $B$) ensures they are energetically identical, protecting the gapless nature of the Dirac points. This same symmetry, however, prevents the opening of a mass gap, which is a prerequisite for defining the distinct, pseudo-chiral states needed to replicate the conditions for the CME. In a strictly $2+1D$ system, the Nielsen-Ninomiya theorem \cite{NIELSEN1983389,Nielsen:1981hk} further constrains the ability to robustly separate opposite-chirality nodes, frustrating the conventional mechanism of the CME.

This dissertation posits that this obstacle can be overcome by moving beyond pristine graphene to engineered honeycomb lattices where the fundamental sublattice symmetry is explicitly broken. By considering a lattice where the two sublattice sites are occupied by different atomic species (such as in monolayer boron nitride), an intrinsic energy difference is introduced. This symmetry breaking opens a band gap and fundamentally alters the low-energy Hamiltonian. In this gapped system, the Hamiltonian acquires a mass term proportional to the Pauli matrix $\sigma_{z}$, a feature that is absent in pristine graphene. Through this association with mass, the $\sigma_{z}$ operator assumes a role directly analogous to the $\gamma^{5}$ chiral operator in $3+1D$ relativistic quantum mechanics. Its eigenvalues correspond to states predominantly localized on one sublattice or the other, creating a ``pseudo-chirality'' that serves as a powerful and direct analogue to the left- and right-handedness of quarks and Weyl fermions.

The central objective of this work is to develop a theoretical model to investigate if a CME-like effect can be generated in such a quasi-planar system. To accomplish this, we will build upon the theoretical framework that identifies a novel form of angular momentum, termed ``lattice spin'' \cite{PhysRevLett.106.116803}, which emerges from the interaction of charge carriers with the underlying honeycomb lattice. This lattice spin is distinct from both conventional electron spin and the dimensionless pseudospin. Our primary research goal is to model how this lattice spin behaves in the presence of an external magnetic field within a system where we have engineered a non-equilibrium imbalance between the pseudo-chiral populations. We hypothesize that the coupling between the magnetic field and the lattice spin in this imbalanced state can induce a net electrical current, realizing a genuine $2+1D$ CME.

\chapter{Chiral Magnetic Effect}
\section{Introduction}\label{chap:introduction}

The Standard Model (SM) of particle physics provides the fundamental framework for describing elementary particles and their interactions. Within it, the theory of the strong force, QCD, details the interactions between quarks and gluons. That force binds protons and neutrons within atomic nuclei, forming the bedrock of matter. While physicists can adeptly calculate high-energy process where quarks and gluons interact weakly — a regime known as the perturbative limit — the deeper, non-perturbative nature of the strong force remains a frontier of active research.

This complexity is starkly apparent in the concept of the QCD vacuum. Far from being empty, the QCD vacuum is a dynamic medium, teeming with quantum fluctuations of both gluon fields and virtual quark-antiquark pairs that constantly appear and disappear. These fields possess a hidden structure, a texture classified by a topological invariant called the winding number, $Q_w$. Just as a ribbon can only have a whole number of twists, these gluon field configurations are characterized by an integer winding number. The existence of configurations with a non-zero $Q_w$ has a profound consequence: in principle, it could break a fundamental symmetry of nature known as Charge-Parity (CP) symmetry. This symmetry posits that the laws of physics should not change if a particle is swapped with its antiparticle (Charge conjugation) while its spatial coordinates are inverted (Parity).  Although, it is a consensus that a global violation of this symmetry in the strong force is not allowed, but a local CP violation by certain QCD field configurations can happen, setting the stage for new and exotic phenomena.

Usually, the quarks and gluons show themselves in nature only in bound states; however, under the extraordinary conditions of extreme temperature and density, like those in the first microseconds after the Big Bang, a new state of matter is predicted to emerge: the QGP. In this primordial soup, quarks and gluons are deconfined. Within the QGP, the topological features of the QCD vacuum are thought to give rise to a remarkable macroscopic phenomenon: the CME \cite{KHARZEEV2008227}. First theorized by Dmitri Kharzeev, Larry McLerran, and Harmen Warringa, the CME describes how, in the presence of an incredibly strong magnetic field, these topological fluctuations can induce a separation of electric charge. Positively charged quarks are pushed along the magnetic field, while negatively charged quarks are pushed in the opposite direction \footnote{This description applies for the case where the topological charges are negative, and for the case where these topological charges are positive, the movement of these charges will be inverse.}. The emergence of this electric current will be explained in detail later.

The place on Earth where such an extreme condition can be recreated is in the fiery crucible of non-central heavy-ion collisions \footnote{Another place where we can find this extreme condition is in compact stars.}. At facilities like the Relativistic Heavy Ion Collider (RHIC) and the Large Hadron Collider (LHC), atomic nuclei are smashed together at nearly the speed of light. The fleeting, off-center collisions of these nuclei are believed to generate some of the most powerful magnetic fields in the known universe, providing the perfect laboratory to search for the CME.

Observing the CME would be a monumental discovery. It would not only provide the first direct experimental evidence of the topological nature of the strong force vacuum but could also imply deep consequences for baryogenesis in the early universe. This exploration delves into the theoretical underpinnings of the CME, explaining the mechanism of its appearance in heavy-ion collision and reviewing the current status of this quest.

\section{Theoretical Framework}\label{chap:theoretical-framework}

The prediction of the CME weaves together three key concepts: the intricate structure of the QCD vacuum, a quantum mechanical effect known as the axial anomaly, and the behavior of quarks in an intense magnetic field.

\subsection{The QCD Vacuum and its Hidden Landscape}\label{subchap:the-qcd-vacuum-and-its-hidden-landscape}

\begin{figure}[h!]
    \centering
    \includegraphics[width=0.8\textwidth]{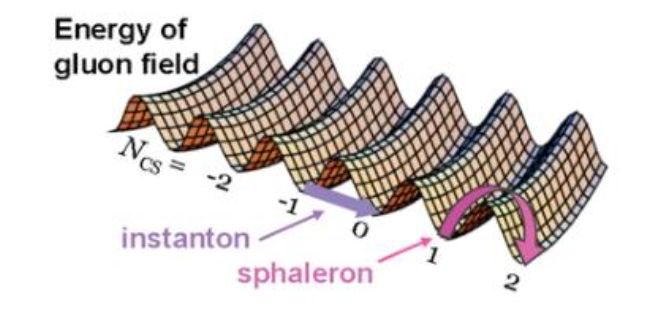}
    \caption{Representation of the QCD energy states.}
    \label{fig:qcd_vaccum}
\end{figure}

Imagine a vast, hilly landscape representing the energy states of QCD. The valleys are the vacuum states — the states of lowest energy. In a simple theory, there might be only one valley. QCD, however, presents a landscape with many distinct valleys, all at the same minimum energy level but separated by hills, or potential energy barriers, separated  by hills and valleys, as we can see in the \autoref{fig:qcd_vaccum}.

These valleys are distinguished by a topological charge, or winding number, an integer characterizing the ``twist'' of the gluon fields in that vacuum state. A transition from one valley to another corresponds to a change in this topological charge, with the total change over time being the winding number $Q_w$. The existence of these distinct vacuum states and the transitions between them is a hallmark of the non-perturbative nature of QCD, and it is these transitions that drive the anomalous processes behind the CME.

\subsection{A Deeper Look at Chirality and Helicity}\label{subchap:a-deeper-look-at-chirality-and-helicity}

Before delving into the quantum anomaly, it is essential to clarify the concepts of chirality and helicity. First, helicity is a physical, observable quantity: it is the projection of a spin particle onto the direction of its momentum. A particle with positive helicity has its spin aligned with its momentum, while one with negative helicity has its spin anti-aligned.

Chirality, on the other hand, is a more abstract property related to how the quantum field of a particle transforms under the Poincaré group. In the mathematical framework of quantum field theory, we can define a chiral operator $\gamma^{5}\approx\gamma^{0}\gamma^{1}\gamma^{2}\gamma^{3}$. This operator allows us to construct projection operators, $P_{R}=\frac{1+\gamma^{5}}{2}$ and $P_{L}=\frac{1-\gamma^{5}}{2}$. When these operators act on the spinor of the particle ($\psi$), they are able to divide it into two distinct and non-overlapping eigenstates: a right-handed component, $\psi_{R}$, and a left-handed component, $\psi_{L}$. The crucial link between these two concepts emerges in the high-energy (or massless) limit. When the energy particle is much greater than its rest mass ($E >> m$), as is the case for quarks in the QGP, the distinction between chirality and helicity effectively vanishes. In this limit, we note that the eigenstates of helicity are also eigenstates of $\gamma^{5}$. Consequently, a right-handed particle ($\psi_{R}$) is one with positive helicity ($\psi_{h\uparrow}$), and a left-handed particle ($\psi_{L}$) is one with negative helicity ($\psi_{h\downarrow}$). This high-energy approximation, where mass is negligible, is very useful in principle to facilitate calculations and provides a direct physical interpretation of chirality.

Classically, the theories of both QED and QCD conserve chirality. This means that in any interaction, the net ``handedness'' is preserved. This has a powerful predictive consequence: if we know the chirality of the particles in an initial state, we can predict the chirality of the final state. Since chirality and helicity are equivalent for $E >> m$, this allows us to predict the final helicity (i.e., the spin-momentum projection) of particles in high-energy collisions.

However, as we will now see, it is precisely this classical conservation law that is violated at the quantum level.

\subsection{Chiral Symmetry and its Quantum Anomaly}\label{subchap:chiral-symmetry-and-its-quantum-anomaly}

To describe the topological effects in the QCD vacuum, we must consider two distinct mechanisms. At low temperatures, they require quantum tunneling through the potential energy barrier, a process described by field configurations called instantons. In high-temperature systems such as QGP, these events are exponentially suppressed ($\sim e^{-\alpha T^2}$; and it will only be suppressed if it is in a thermal medium of QGP), making them infrequent.

However, in the scorching environment of a QGP, the immense thermal energy allows the system to simply pass classically over the potential barriers. These high-temperature transitions are mediated by unstable, static field configurations called sphalerons, from the Greek for ``ready to fall''. Unlike instanton-driven tunneling, sphaleron transitions are not suppressed and are expected to occur at a high rate in the QGP.

The primary consequence of these rapid transitions is their effect on a core principle of the theory known as chiral symmetry. In a simplified world of massless quarks, QCD equations have this special symmetry. The ``chirality'' refers to the ``handedness'' of a particle. Quarks can be left-handed or right-handed. In the limit of massless particles, these quantities are identical to helicity; left-handed implies spin opposite to momentum, and right-handed implies spin aligned to momentum. Chiral symmetry implies that the numbers of left-handed and right-handed quarks are each separately conserved in classical interactions.

However, this classical symmetry is broken at the quantum level by the axial anomaly. This is a profound effect where the classical symmetry is violated when the theory is quantized. The axial anomaly directly links the topological transitions in the QCD vacuum to a change in the net chirality of the system.

Specifically, when a gluon field configuration has a non-vanishing topological charge ($Q_w \neq 0$), the axial current is no longer conserved (even for the massless quarks). This means that the difference between the number of left-handed ($N_L$) and right-handed ($N_R$) quarks changes according to the relation

\begin{equation}
    (N_L - N_R) = 2N_f Q_w,
\end{equation}

\noindent where $N_f$ is the number of quark flavors. A gluon configuration with a positive $Q_w$ interacts with quarks that convert right-handed into left-handed ones, while a negative $Q_w$ does the opposite. This anomalous chirality-flipping process is the first crucial ingredient for the CME, providing the mechanism to create a temporary, localized imbalance between left- and right-handed quarks in the QGP.

This ensures that local domains with a net chirality imbalance — an excess of left- or right-handed quarks — are frequently and randomly produced throughout the plasma lifetime.

\subsection{The Chiral Magnetic Effect}

The CME emerges when a local imbalance in chirality, created by sphaleron transitions, combines with a powerful external magnetic field estimate in $10^{19}$ Gauss.

\subsubsection{An Idealized Picture}

\begin{figure}[h!]
    \centering
    \includegraphics[width=0.6\textwidth]{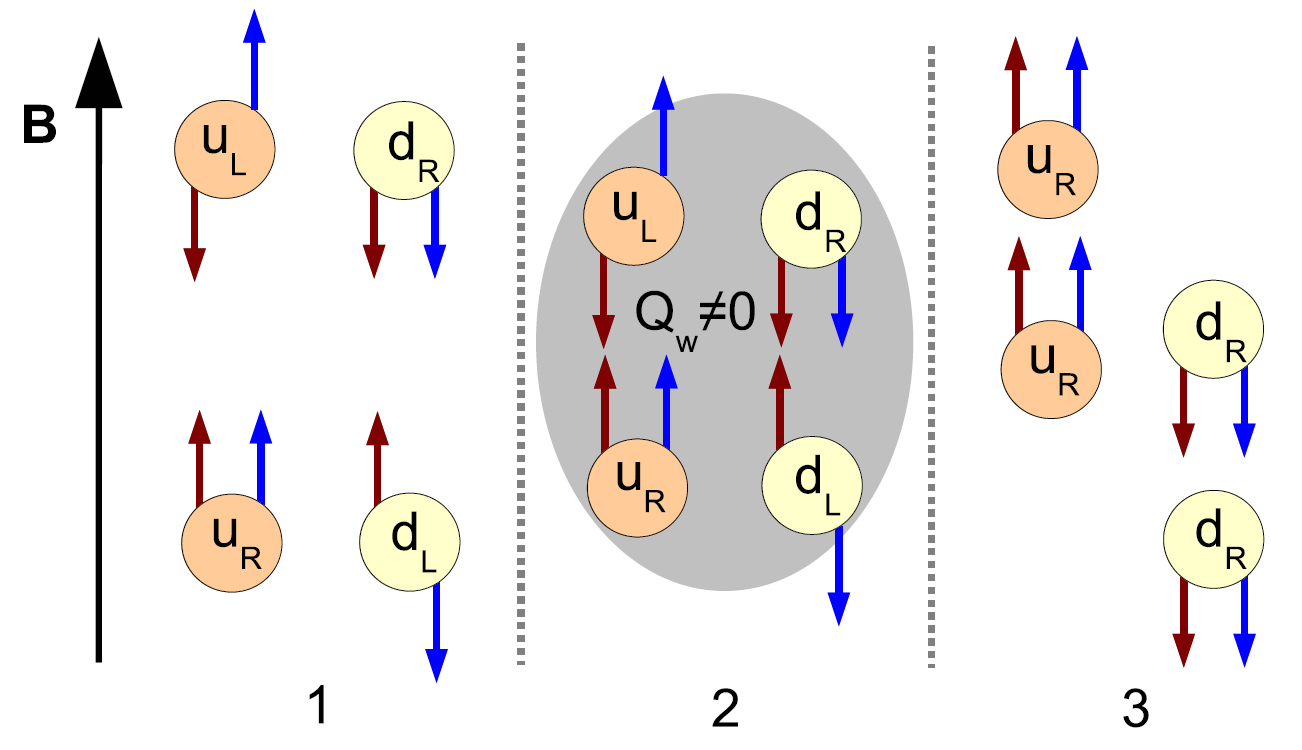}
    \caption{An illustration of the CME in a very large, homogeneous magnetic field. The red arrows denote momentum, and the blue arrows denote spin. The figure is broken down into three stages, read from left to right. Figure extracted from the reference \cite{KHARZEEV2008227}.}
    \label{fig:cme}
\end{figure}

To understand the mechanism, consider a small region in the QGP where a sphaleron transition has created an excess of right-handed quarks. Now, immerse this region in an extremely strong magnetic field ($\mathbf{B}$). The process is illustrated in \autoref{fig:cme}.

\begin{itemize}
    \item \textbf{Initial State}: On the left, we see the initial state. Due to the very large magnetic field pointing upward, all quarks (up and down) are in their lowest Landau level. This forces their spins (blue arrows) to align with the magnetic field and restricts their movement to be parallel to the field lines. Positively charged up quarks ($u$) align their spin with $\mathbf{B}$, while negatively charged down quarks ($d$) align their spin opposite to $\mathbf{B}$. Initially, there are equal numbers of left-handed ($u_L, d_L$) and right-handed ($u_R, d_R$) quarks, and as a consequence, we have these quarks up and down moving in both directions.

    \item \textbf{Topological Transition}: In the center, the quarks interact with a gauge field configuration having a non-zero topological winding number, $Q_w \neq 0$. In the example shown, $Q_w=-1$, which converts left-handed up and down quarks into right-handed ones. In such a strong magnetic field, this change in helicity is achieved by reversing the momentum of the quarks (red arrows), not by flipping their spins, which is energetically suppressed.

    \item \textbf{Charge Separation}: On the right, we see the final state. There is now an excess of right-handed quarks. The right-handed up quarks ($u_R$) move upwards, parallel to the magnetic field. The right-handed down quarks ($d_R$) also move in the direction of their spin, which is oriented opposite to the magnetic field, so they move downward. This results in a separation of charges: positive charges flow up, and negative charges flow down. This separation of charge is the CME, which effectively induces an electric current along the direction of the magnetic field.
\end{itemize}

\subsection{The Real-World Scenario}

In a heavy-ion collision, the magnetic field is not infinite. This means that the spin alignment of the quarks is not perfect; the degree of polarization depends on the strength of the field. Consequently, the charge separation is smaller than in the ideal case. The magnitude of the effect depends on the magnetic field strength, the plasma temperature, and the sphaleron transition rate.

\section{The Search in Heavy-Ion Collisions}

Non-central heavy-ion collisions provide the only known terrestrial setting where the necessary conditions for the CME can be realized. The experimental search for the CME in the QGP is a major focus at facilities like the RHIC and the LHC, as its definitive observation would provide profound evidence for the existence of topological charge fluctuations in QCD.

When large nuclei (e.g., gold) collide off-center at relativistic speeds, the protons that do not directly hit (the ``spectators'') continue on their path. These fast-moving positive charges create a transient but immense magnetic field, estimated \cite{KHARZEEV2008227} to be on the order of $10^2-10^3 \, \text{MeV}^2$ (which corresponds to roughly $10^{19}$ Gauss), trillions of times stronger than any lab-created field. This massive magnetic field is directed perpendicular to the reaction plane — the plane defined by the beam direction and the impact of the collision parameter. Direct collision of the other nucleons (the ``participants'') creates the QGP, the deconfined and chirally restored medium, where frequent sphaleron transitions can generate local domains of chirality imbalance. Both participants and spectators are jointly responsible for inducing the magnetic field; it is not a product of the spectators in isolation.

\subsection{From a Fleeting Current to a Measurable Signal}

An electric current deep within the dense QGP would likely dissipate quickly due to constant rescattering. Therefore, the observable effect of the CME is thought to be dominated by transitions near the surface of the plasma. If a sphaleron transition near the edge pushes positive quarks outward and negative quarks inward, the outgoing positive quarks can escape and be detected. However, the inward-moving negative quarks will likely be reabsorbed and scattered, losing their initial correlation with the magnetic field direction. This surface effect would lead to a net observable asymmetry in the emitted particles.

\subsection{The Experimental Signature}

A crucial feature of sphaleron transitions is that their topological charge ($Q_w = \pm 1$) fluctuates randomly from event to event. In some events, positive charges separate in one direction; in others, they separate in the opposite direction. Averaging over many events would yield a net effect of zero, making the CME seemingly invisible.

To circumvent this, physicists look for the variance of charge separation using correlation techniques. Instead of looking for a net charge separation, they analyzed correlations between the emitted charged particles relative to the reaction plane. The CME predicts that particles with the same electric charge should be preferentially emitted on the same side of the reaction plane. In contrast, particles with opposite charges should be preferentially emitted on opposite sides.

Experiments at RHIC and LHC measure the angles of emitted charged particles and calculate correlators sensitive to this pattern. A positive correlation for same-sign pairs and a negative one for opposite-sign pairs are the expected signatures of the CME. The signal strength is predicted to depend on factors like collision energy, the size and charge ($Z$) of the nuclei, and the centrality collisions, which determine the magnetic field strength. The search for this subtle effect continues to be a major focus of high-energy nuclear physics, promising a deeper understanding of the fundamental laws governing our universe. 

\begin{figure}[h!]
    \centering
    \includegraphics[width=0.6\textwidth]{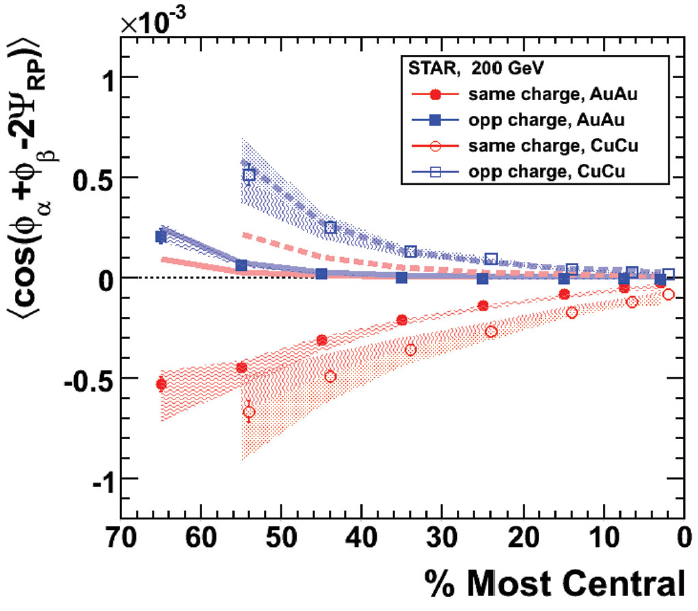}
    \caption{The measured three-particle correlator $\langle\cos(\phi_{\alpha}+\phi_{\beta}-2\Psi_{RP})\rangle$ as a function of collision centrality for $Au+Au$ and $Cu+Cu$ systems at 200 GeV. Same-charge pairs are shown with circles (red) and opposite-charge pairs with squares (blue). Figure extracted from the reference \cite{PhysRevLett.103.251601}.}
    \label{fig:cme_experiment}
\end{figure}

In 2009, the STAR Collaboration at RHIC published \cite{PhysRevLett.103.251601} findings from $Au+Au$ and $Cu+Cu$ collisions that seemed to fit this exact pattern. Using a three-particle azimuthal correlator, $\langle\cos(\phi_{\alpha}+\phi_{\beta}-2\Psi_{RP})
\rangle$, which is directly sensitive to charge separation, they observed a distinct difference between same-charge and opposite-charge particle correlations. The results, summarized in \autoref{fig:cme_experiment} from their publication, show several features consistent with the predictions for the CME.

This figure is central to the argument of the paper, as it visually represents the key experimental evidence. Here is how it endorses the idea of local parity violation:

\begin{itemize}
    \item \textbf{Clear Separation of Charges}: The most striking feature is that the correlations for same-charge pairs (circles) are negative, while the correlations for opposite-charge pairs (squares) are positive. This separation is a primary prediction of the CME, where the same-sign charges are pushed in one direction relative to the reaction plane and the opposite-sign charges in the other.
    
    \item \textbf{Centrality Dependence}: Moving from left to right on the x-axis, the collisions become more central (i.e., more head-on and producing more particles). The magnitude of the correlation signal for both the same and opposite charges decreases as the centrality increases. This is consistent with theoretical expectations that the CME signal should be inversely proportional to the number of charged particles produced.
    
    \item \textbf{System-Size Dependence}: The signal in $Cu+Cu$ collisions is consistently larger than in $Au+Au$ collisions at the same centrality. This observation supports a theoretical prediction that the suppression of back-to-back charge correlations (which affects the opposite-charge signal) would be smaller in a lighter collision system like $Cu+Cu$.
    
    \item \textbf{Magnitude Agreement}: In the $Au+Au$ system, the opposite-charge correlations are clearly smaller in magnitude than the same-charge correlations. This aligns with the theoretical possibility that interactions within the hot, dense medium created in the collision suppress back-to-back correlations.
\end{itemize}

The STAR team carefully considered various background sources unrelated to parity violation but found that conventional event-generator models such as HIJING \cite{PhysRevD.44.3501}, URQMD \cite{MBleicher_1999}, and MEVSIM \cite{ray2000mevsimmontecarloevent} could not adequately explain the observed same-charge correlations. This led to the conclusion that the signal was consistent with the formation of parity-odd domains, providing tantalizing evidence for the first observation of local strong-parity violation.

However, a year later, a theoretical paper by Scott Pratt \cite{pratt2010alternativecontributionsangularcorrelations} presented a compelling counterargument. Pratt proposed that the observed correlations might not be a signature of new physics but rather the product of conventional physical principles that had not been fully appreciated as a background. He argued that the bulk of the signal could be explained by combining two well-understood phenomena:

\begin{itemize}
    \item \textbf{Local Charge Conservation}: When a positive-negative particle pair is created, they originate from the same point, and the collective elliptic flow of the medium tends to push them in a similar direction, naturally creating a correlation for opposite-sign pairs.
    \item \textbf{Momentum Conservation}: The overall conservation of momentum, when combined with the anisotropic nature of elliptic flow, inherently generates a back-to-back correlation that strongly affects same-sign pairs.
\end{itemize}

Pratt demonstrated that these effects could produce correlations with magnitudes similar to those measured by the STAR. He contended that what was being interpreted as a signal for the CME could be largely a consequence of this interplay between conservation laws and collective flow. His work suggested that this ``background'' is so significant that it could account for half or more of the entire observed effect.

This has led to a crucial scientific debate. The STAR collaboration results remain consistent with the CME, but the work of Pratt provides a substantial, quantifiable alternative explanation rooted in known physics. His paper does not disprove the existence of the CME but rather establishes a critical baseline; any true signal from parity violation must be shown to exist above and beyond the correlations generated by conservation laws and flow. To distinguish between the two scenarios, Pratt suggested more differential measurements, such as analyzing the correlations as a function of the particle separation in rapidity. Correlations from conservation laws are expected to be short-range, confined to about one unit of relative rapidity, whereas effects from the CME are predicted to extend over a much wider range. This ongoing dialogue highlights the complexity of untangling new phenomena from the intricate backdrop of known physics in heavy-ion collisions.

To address this complex scientific debate and better isolate the potential CME signal from flow-related backgrounds, the STAR Collaboration designed a decisive experiment \cite{PhysRevC.105.014901} using collisions of isobar nuclei: Ruthenium-96 ($_{44}^{96}\mathrm{Ru}$) and Zirconium-96 ($_{40}^{96}\mathrm{Zr}$). These two nuclei have the same mass number (96), meaning that they should produce nearly identical backgrounds related to collective flow and particle multiplicity. However, Ruthenium has four more protons than Zirconium ($Z=44$ vs. $Z=40$), resulting in a magnetic field that is predicted to be roughly $10-15\%$ stronger in Ru+Ru collisions. Therefore, if the CME is real, it should produce a measurably larger charge-separation signal in $Ru+Ru$ collisions compared to $Zr+Zr$ collisions, while the background remains the same.

\begin{figure}[h!]
    \centering
    \includegraphics[width=0.99\textwidth]{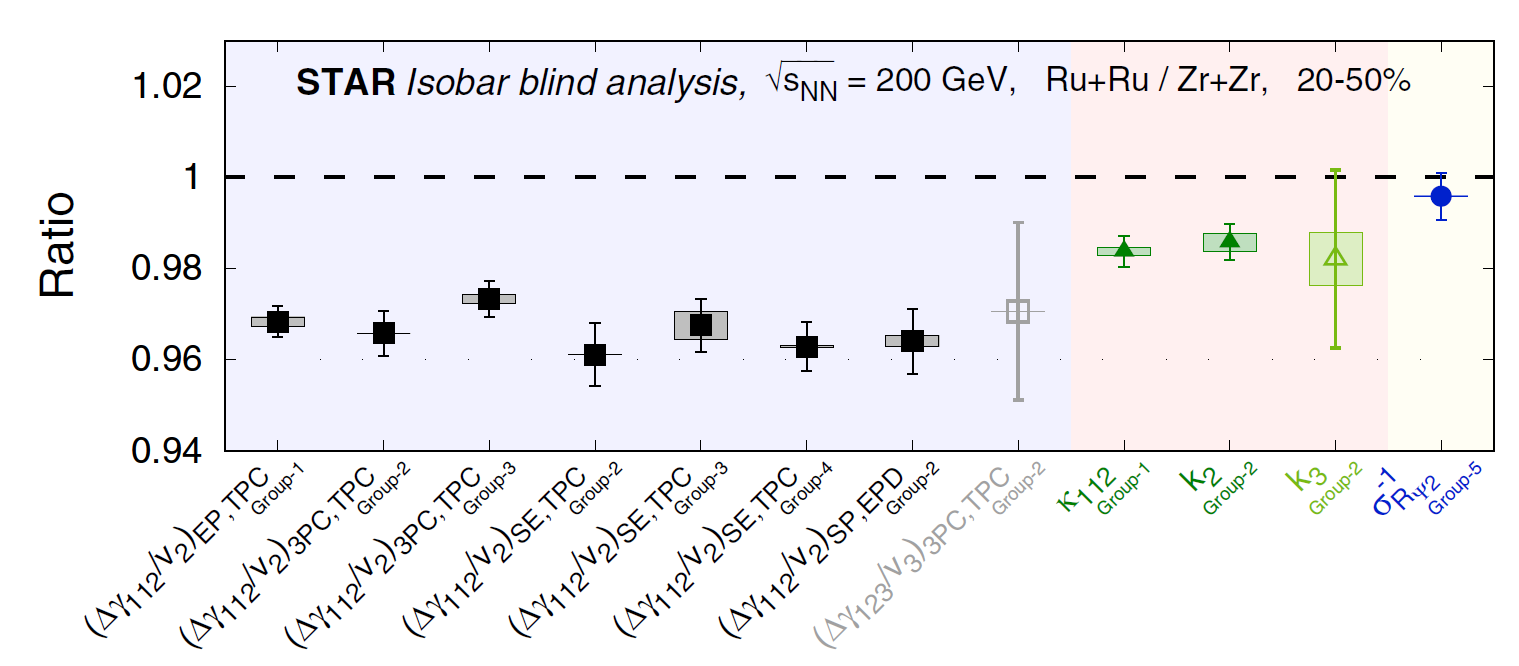}
    \caption{No predefined signatures of the CME were observed in this blind analysis. The study compared results from Ruthenium ($Ru+Ru$) and Zirconium ($Zr+Zr$) collisions. It found that the ratio of CME-sensitive measurements was below one, leading to the conclusion of a null result. The presented data distinguishes between CME-sensitive (solid symbols) and insensitive (open symbols) measures, showing both statistical and systematic uncertainties. Figure extracted from the reference \cite{PhysRevC.105.014901}.}
    \label{fig:isobar}
\end{figure}

In 2022, the STAR Collaboration published the results of a blind analysis of $3.8$ billion of these isobar collisions. The findings, summarized in \autoref{fig:isobar} from the publication, did not show evidence of a CME signal that met the predefined criteria. The experiment achieved a remarkable precision of $0.4\%$ in comparing the CME-sensitive observables between the two systems. As shown in \autoref{fig:isobar}, the ratio of the key observable $\Delta
\gamma/v_{2}$ for $Ru+Ru$ over $Zr+Zr$ was consistently found to be less than one, contrary to the expectation of a value greater than one for a CME signal. Other CME-sensitive observables also did not show enhancement in Ruthenium collisions. The results indicated that differences in the nuclear structure and multiplicity between the two isobars, rather than the CME, were the dominant factors. This landmark study, while not entirely ruling out the existence of the CME, placed stringent constraints on its possible magnitude and concluded that no definitive signal had been observed. The expected result was a value greater than or equal to one ($\geq1$). However, the result obtained did not meet this expectation, suggesting that the data was not processed correctly. We conclude that after a thorough correction of the data treatment, we expect to observe different results that align with the initial hypothesis.

\section{Bridge between QCD and materials}

As discussed before, in the context of high-energy heavy-ion collisions, it is theorized that localized domains can form within the QGP where fundamental symmetries like parity (P) and charge-parity (CP) are violated. These domains are characterized by a nonzero topological charge which, through the chiral anomaly, manifests as an imbalance in the number of left-handed ($N_L$) and right-handed ($N_R$) quarks. This imbalance is quantified by a chiral chemical potential, $\mu_{5} = \mu_{R}-\mu_{L}\neq 0$. In the presence of the immensely strong magnetic fields ($\mathbf{B}\sim 10^{14}T$) generated transiently in non-central heavy-ion collisions, this chirality imbalance is predicted to induce a separation of electric charge, resulting in a net electric current flowing along the magnetic field axis. The CME current density ($\mathbf{J}_{CME}$) \cite{Li2016} is elegantly described by the formula,

\begin{equation}
    \mathbf{J}_{CME} = \frac{e^2}{2\pi^2}\mu_5\mathbf{B}. 
\end{equation}

This equation encapsulates a remarkable phenomenon: a dissipation-less current driven by a quantum anomaly. A non-dissipative current is one that arises from a process that does not break time-reversal symmetry, making it intrinsically dissipation-less with no associated change in entropy. This effect generates an electric current through an imbalance between left-handed and right-handed fermions within a steady magnetic field, unlike conventional electricity which requires a changing magnetic field. This corresponds to a current aligned with the direction of the field, a phenomenon which is not predicted by the standard equations of Maxwell \footnote{On the presence of the chiral anomaly the equations of Maxwell receive new terms.}. This presents a potentially revolutionary way to transfer charge without resistance. 

Conservation Laws and the Anomaly
Underpinning the entire discussion of the CME is the concept of current conservation and the role of quantum anomalies. In classical field theory, both the vector current (associated with electric charge) and the axial current (associated with chirality) are conserved quantities. However, quantum mechanics introduces a profound subtlety: the chiral anomaly. The anomaly dictates that while the vector current remains conserved, the axial current is not. The chiral anomaly manifests in both abelian and non-abelian forms, each arising from different physical mechanisms. The non-abelian anomaly, as established earlier, is linked to topological field configurations like instantons and sphalerons. Conversely, the abelian anomaly arises from the presence of parallel electric and magnetic fields, and its mathematical expression is given by \autoref{non-conservation_of_axial_charge}.

\begin{equation}
    \partial_\mu\mathbf{J}^{\mu}_{5} = \frac{e^2}{16\pi^2}F_{\mu\nu}\tilde{F}^{\mu\nu}\propto\mathbf{E}\cdot\mathbf{B}.
    \label{non-conservation_of_axial_charge}
\end{equation}

This equation is the heart of the matter. It states that in the presence of parallel electric and magnetic fields, the chiral charge is not conserved. It is precisely this ``violation'' of classical conservation that allows the pumping of charge between Weyl nodes and the generation of a chiral chemical potential $\mu_5$.

The key insight that translated the CME to condensed matter physics was the recognition that the fundamental ingredients — massless chiral fermions and the potential for a chiral imbalance — are not exclusive to the QGP. In the late 2000s, it was realized that a class of newly discovered topological materials, known as Dirac and Weyl semimetals \cite{PhysRevB.83.205101,PhysRevLett.107.127205,PhysRevB.86.115133}, provide a natural solid-state arena for this effect. These materials are characterized by a unique electronic band structure where the conduction and valence bands touch at discrete points in momentum space, called Dirac or Weyl nodes. Near these nodes, the low-energy electronic excitations behave as relativistic massless Dirac or Weyl fermions, governed by a linear energy-momentum dispersion relation.

Weyl semimetals are particularly interesting because their electronic structure hosts unique points called Weyl nodes, where electrons behave like massless particles. These nodes always come in pairs of opposite ``handedness'', or chirality. The separation of these nodes in the momentum space, $\Delta k$, effectively acts as an axial gauge field. When an external electric field ($\mathbf{E}$) is applied parallel to an external magnetic field ($\mathbf{B}$), a quantum pumping of charge occurs between nodes of opposite chirality, a process known as the Adler-Bell-Jackiw (ABJ) chiral anomaly. This pumping creates a chiral imbalance ($\mu_5\neq 0$), fulfilling the necessary condition for the CME. 

The following equation describes the CME current by relating it to the electric field $\mathbf{E}$ through the CME conductivity tensor $\sigma^{ik}_{CME}$ \cite{Li2016},

\begin{equation}
    \mathbf{J}^{i}_{CME} = \frac{e^{2}}{\pi\hbar}\frac{3}{8}\frac{e^{2}}{\hbar c}\frac{v^{3}}{\pi^{3}}\frac{\tau_{V}}{T^{2}+\frac{\mu^{2}}{\pi^{2}}}\mathbf{B}^{i}\mathbf{B}^{k}\mathbf{E}^{k}\equiv\sigma^{ik}_{CME}\mathbf{E}^{k},
\end{equation}

\noindent where in this expression, $\mathbf{J}_{CME}^{i}$ is the anomalous current, driven by an external electric field ($\mathbf{E}^{k}$) and generated through the interplay of several factors. These include fundamental constants like the elementary charge ($e$), the reduced Planck constant ($\hbar$), and the speed of light ($c$), alongside material-specific properties. Key among these are the Fermi velocity ($v$) of the relativistic-like quasi-particles, the absolute temperature of the system ($T$), and its chemical potential ($\mu$). A critical component is the chirality-changing scattering time ($\tau_{V}$), which represents the average time it takes for a quasi-particle ``handedness'' to be flipped by interactions within the material. The dependence of the effect on the external magnetic field ($\mathbf{B}^{i}$, $\mathbf{B}^{k}$) is quadratic, which is a primary experimental signature of the CME. This entire response is captured by the CME conductivity tensor, $\sigma_{CME}^{ik}$.

When the electric and magnetic fields are parallel, the CME conductivity tensor \cite{Li2016} simplifies to a single longitudinal component, $\sigma_{CME}^{zz}$, as shown in \autoref{eq:CME-conductivity-tensor}. This equation, 

\begin{equation} \label{eq:CME-conductivity-tensor}
    \sigma^{zz}_{CME} = \frac{e^{2}}{\pi\hbar}\frac{3}{8}\frac{e^{2}}{\hbar c}\frac{v^{3}}{\pi^{3}}\frac{\tau_{V}}{T^{2}+\frac{\mu^{2}}{\pi^{2}}}\mathbf{B}^{2},
\end{equation}

\noindent isolates the response of the material under the precise conditions required to observe the CME. The conductivity depends on a combination of fundamental constants (the elementary charge ($e$), the reduced Planck constant ($\hbar$), and the speed of light ($c$) and material-specific properties, such as the Fermi velocity ($v$) of the quasi-particles, the temperature of the system ($T$), and its chemical potential ($\mu$). The crucial term $\tau_{V}$ represents the chirality-changing scattering time, which quantifies the rate at which quasi-particles flip their ``handedness'' due to interactions within the material. 

The most importantly, the equation explicitly shows that the conductivity is proportional to the square of the magnetic field strength ($\mathbf{B}^2$), a distinctive quadratic dependence that serves as a primary and unambiguous experimental signature of the CME.

So, in this case, in addition to the usual role that $\mathbf{B}$ already had, it is also responsible for the chirality imbalance. Consequently, applying parallel electric and magnetic fields to a Weyl semimetal is predicted to generate a substantial anomalous contribution to the electric current, leading to a large, negative longitudinal magnetoresistance (LMR) — a key experimental signature.

\begin{figure}[h!]
    \centering
    \includegraphics[width=0.6\textwidth]{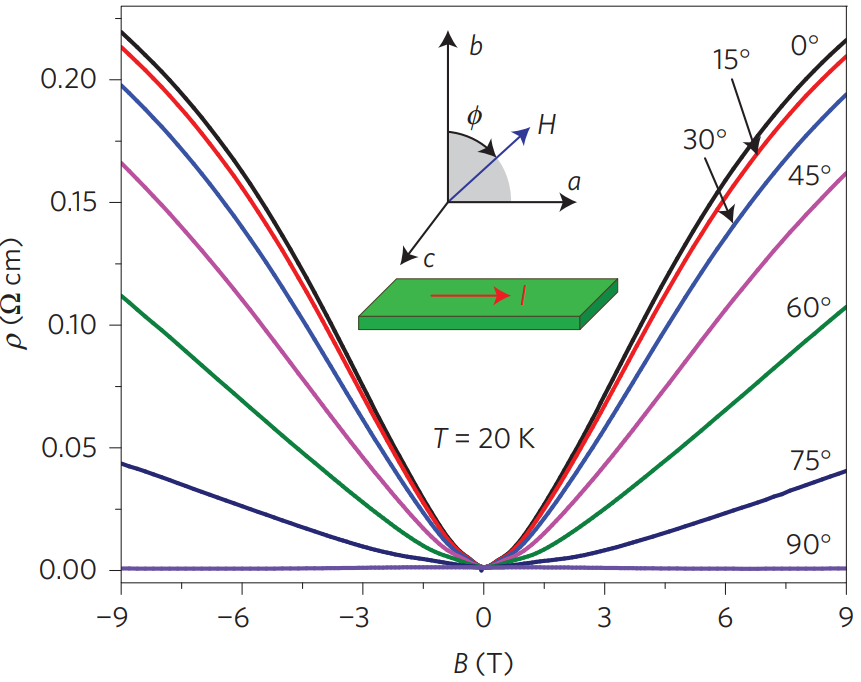}
    \caption{This graph is critically important as it shows that when the magnetic field is aligned parallel to the current (at an angle of $90 ^{\circ}$), the material exhibits a large negative magnetoresistance. This specific phenomenon — a decrease in resistance when electric and magnetic fields are parallel — is a key signature of the CME. Figure extracted from reference \cite{Li2016}.}
    \label{fig:chiral_magnetic_effect_in_ZrTe5}
\end{figure}

In 2013, the theoretical predictions were stunningly confirmed. Experiments on materials such as $ZrTe_{5}$ \cite{Li2016} (a Dirac semimetal) and later $TaAs$ \cite{PhysRevX.5.031013} and $Na_{3}Bi$ \cite{PhysRevX.5.031023} (Weyl semimetals) reported the observation of a large negative LMR when the electric and magnetic fields were aligned. This effect, which persisted at relatively high temperatures and was robust against scattering, was accompanied by a characteristic quadratic dependence on the magnetic field [\autoref{fig:chiral_magnetic_effect_in_ZrTe5}], providing strong evidence for its anomalous, CME - driven origin. 

These discoveries were a landmark achievement, demonstrating that the physics of the chiral anomaly, once confined to the abstract realm of high-energy theory, could be harnessed and measured in a $3+1 D$ solid-state system.

The successful observation of the CME in $3+1 D$ materials naturally poses the next fundamental question: can this effect manifest in systems of lower dimensionality? The focus of this inquiry inevitably turns to the quintessential $2+1 D$ material, graphene. The electronic graphene structure is celebrated for its two inequivalent valleys (considering in terms of chirality [subsection \ref{subchap:chirality-and-pseudospin}]) $K^{+}$ and $K^{-}$, where electrons behave as massless Dirac fermions. This immediately suggests a parallel with the chiral fermions required for the CME, with the valley index playing a role analogous to chirality. Therefore, inducing a valley imbalance in graphene — a ``chiral chemical potential'' in this context — in the presence of a perpendicular magnetic field might be expected to produce a CME, like current.

However, the transition from $3+1 D$ to $2+1 D$ presents significant theoretical challenges. The Nielsen-Ninomiya theorem, a cornerstone of topological band theory, dictates that chiral fermions in a lattice system must appear in pairs of opposite chirality. In a strictly $2+1 D$ system, it is generally impossible to separate these opposite-chirality nodes in momentum space in a way that allows for a net chiral charge to be robustly defined and manipulated as it is in $3D$ Weyl semimetals. The CME in its conventional form is intrinsically a three-dimensional phenomenon, requiring particle momentum to have a component parallel to the magnetic field. In a perfectly planar $2D$ system, the electron momenta are confined to the plane, and a magnetic field applied perpendicular to this plane cannot induce the same kind of chiral charge separation. The very mechanism of the CME seems to be frustrated by the reduction in dimensionality.

This raises a crucial and subtle point: what constitutes a truly $2+1 D$ system? Real-world graphene is never perfectly planar. It exhibits intrinsic ripples, corrugations, and can be supported on substrates that induce structural variations. Furthermore, from a quantum mechanical perspective, the electron wavefunction is not infinitely confined to a $2D$ plane but has a finite extent in the third dimension, on the order of the atomic height. The central question we wish to explore is whether this minimal, often-neglected dimension is sufficient to break the strict $2+1 D$ prohibition and allow a CME - like effect to emerge.

If the height of a single atom, or the nanoscale curvature of a ripple, is enough to provide a finite z-component for electron motion, then a magnetic field with an in-plane component could, in principle, drive the CME. Strain, curvature, and substrate interactions in graphene can act as pseudo-gauge fields, potentially mimicking the effects of axial fields needed to generating an analogy to the chiral imbalance, having this role played by a unbalance between the gap of the Dirac cones in the material. For instance, a spatially varying strain field could create a pseudo-magnetic field that is oppositely directed in the $K^{+}$ and $K^{-}$ valleys, effectively creating a valley polarization. If a real magnetic field is then applied, the conditions for the CME could be met, albeit in a more complex, spatially dependent manner. The current would not be a bulk effect as in a $3D$ semimetal but might manifest as edge currents or along specific strain-induced channels. The challenge lies in engineering and detecting such subtle effects, which are likely to be much smaller than their $3D$ counterparts. While this effect cannot be observed in pristine graphene, it can be search in materials that possess a similar structure and electronic properties.

The CME is the physical consequence of the attempt of the system to respond to the anomaly-induced chiral imbalance. The conservation of the vector current, however, remains sacrosanct. The current generated by the CME is a real physical electric current.\vphantom{The anomaly does not create an electric charge from nothing; it merely converts a topological property of the gauge fields into a measurable charge separation.} Understanding this interplay — the non-conservation of the axial current leading to a physical, conserved vector current — is essential for any theoretical or experimental investigation of the CME, regardless of the dimensionality of the system.

We will now proceed with a detailed examination of the structure and symmetries of graphene and related materials. This analysis is essential for understanding the properties characteristic of the materials discussed later in this work.

\chapter{Graphene and Symmetries}
The preceding discussion established a  bridge that connects high-energy physics with observable phenomena in three-dimensional topological materials. The successful observation of the CME in Weyl and Dirac semimetals naturally leads to the next frontier: Can this fundamentally $3D$ effect be realized in a $2+1D$ system? As was argued, graphene, with its unique electronic structure hosting massless Dirac fermions, stands as the primary candidate for this investigation.

However, answering this question is not straightforward and requires a deep understanding of the intrinsic properties of the material. This chapter is therefore dedicated to a detailed examination of the graphene lattice structure and its crucial symmetries. By dissecting the very features that grant graphene its celebrated electronic character — the hexagonal lattice, the emergence of Dirac cones, and the valley degree of freedom that acts as a form of pseudo-chirality — we lay the essential groundwork. This analysis is not merely descriptive; it is the necessary first step in constructing a theoretical framework to understand if and how the subtle interplay of chirality and magnetism, central to the CME, can manifest in the planar world of graphene.

Although pristine graphene is a material with remarkable electronic and structural properties, it is not directly applicable to the objectives of this study because it lacks the specific functionalities required. However, its fundamental structure serves as an essential starting point. This chapter, therefore, focuses on using graphene as a foundational platform for the development and analysis of more complex derivative materials that possess the necessary characteristics for our proposed application.

\section{The Honeycomb Lattice and Reciprocal Space}

In this section, we are going to understand the characteristics and properties of graphene following the lines used in \cite{graphene}. 

Graphene has a distinctive honeycomb crystal lattice, which is not a Bravais lattice (for a review of the Bravais lattice, see \autoref{appendix:a}) itself, but rather a triangular Bravais lattice with a two-atom basis. These two atoms belong to different sublattices, conventionally labeled $A$ and $B$, such that each $A$ atom is exclusively bonded to $B$ atoms, and vice versa.

 \begin{figure}[htb!]
     \centering
     \includegraphics[width=0.4\linewidth]{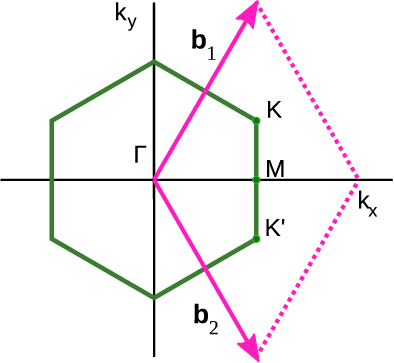}
     \caption{The reciprocal lattice of graphene.}
     \label{fig:graphene_reciprocal-lattice}
 \end{figure}

The lattice vectors for the triangular Bravais lattice are given by

\begin{equation}
    \vec{a}_{1} = \frac{a}{2}(3,\sqrt{3}), \quad \vec{a}_{2} = \frac{a}{2}(3,-\sqrt{3}),
\end{equation}

\noindent where $a \approx 1.42 \, \text{\AA}$ is the nearest-neighbor carbon-carbon distance.

The vectors connecting an atom to its three nearest neighbors are

\begin{equation}
    \vec{\delta}_{1} = \frac{a}{2}(1,\sqrt{3}), \quad \vec{\delta}_{2} = \frac{a}{2}(1,-\sqrt{3}), \quad \vec{\delta}_{3} = a(-1,0).
    \label{eq:vectors-nearest-neighbors}
\end{equation}

In reciprocal space, the lattice is also triangular, with reciprocal lattice vectors (for a review of reciprocal lattice, see \autoref{appendix:a}) represented by

\begin{equation}
    \vec{b}_{1} = \frac{2\pi}{3a}(1,\sqrt{3}), \quad \vec{b}_{2} = \frac{2\pi}{3a}(1,-\sqrt{3}).
\end{equation}

The first Brillouin zone is hexagonal (see \autoref{fig:graphene_reciprocal-lattice}). Of particular importance are the high-symmetry points at the corners of the Brillouin zone, known as the $K$ and $K'$ points (or Dirac points). Their coordinates are

\begin{equation}
    \vec{K}' = \left(\frac{2\pi}{3a},\frac{2\pi}{3\sqrt{3}a}\right), \quad \vec{K} = \left(\frac{2\pi}{3a},-\frac{2\pi}{3\sqrt{3}a}\right).
    \label{eq:k-points}
\end{equation}

Another high-symmetry point is the point $M$, located at the midpoint of a Brillouin zone edge

\begin{equation}
    \vec{M} = \left(\frac{2\pi}{3a},0\right).
\end{equation}

\section{Tight-Binding Model for $\pi$ States}

The electronic band structure of graphene can be effectively described using a tight-binding model, focusing initially on the $\pi$ states derived from the $p_z$ orbitals and considering only nearest-neighbor interactions. Since the honeycomb lattice has two atoms ($A$ and $B$) per unit cell, the basis consists of two states, $|\psi_A\rangle$ and $|\psi_B\rangle$. Hopping is assumed to occur only between atoms on different sublattices ($A$ to $B$ or $B$ to $A$) with a hopping parameter $t$.

The Hamiltonian for a given wave vector $\vec{k}$ in this approximation is represented by a $2\times2$ matrix

\begin{equation}
    \hat{H}(\vec{k}) = \begin{pmatrix}
        0 & tS(\vec{k}) \\
        tS^{*}(\vec{k}) & 0
    \end{pmatrix}.
\end{equation}

Here, the diagonal elements are zero because there is no hopping within the same sublattice. The off-diagonal elements represent the hopping between sublattices. The term $S(\vec{k})$ sums the phase factors acquired by an electron hopping to the nearest neighbors 

\begin{equation}
    S(\vec{k}) = \sum_{\vec{\delta}}e^{i\vec{k}\cdot\vec{\delta}}.
\end{equation}

Using the nearest-neighbor vectors $\vec{\delta}_{1,2,3}$, defined in \autoref{eq:vectors-nearest-neighbors}, this sum is evaluated to 

\begin{equation}
    \begin{split}
         S(\vec{k}) = \sum_{\vec{\delta}}e^{i\vec{k}\cdot\vec{\delta}} &= e^{i\vec{k}\cdot\vec{\delta}_1} + e^{i\vec{k}\cdot\vec{\delta}_2} + e^{i\vec{k}\cdot\vec{\delta}_3} \\
         & = e^{i(k_{x}\frac{a}{2}+k_{y}\frac{a\sqrt{3}}{2})} + e^{i(k_{x}\frac{a}{2}-k_{y}\frac{a\sqrt{3}}{2})} + e^{i(-k_{x}a)} \\
         & = \left[\cos{\left(k_{x}\frac{a}{2}\right)}+i\sin{\left(k_{x}\frac{a}{2}\right)}\right]\cdot\left[\cos{\left(k_{x}\frac{a\sqrt{3}}{2}\right)}+i\sin{\left(k_{y}\frac{a\sqrt{3}}{2}\right)}\right] \\
         & + \left[\cos{\left(k_{x}\frac{a}{2}\right)}+i\sin{\left(k_{x}\frac{a}{2}\right)}\right]\cdot\left[\cos{\left(k_{x}\frac{a\sqrt{3}}{2}\right)}-i\sin{\left(k_{y}\frac{a\sqrt{3}}{2}\right)}\right] \\
         & + e^{i(-k_{x}a)} \\
         & = 2\exp\left(\frac{ik_{x}a}{2}\right)\cos{\left(\frac{k_{y}a\sqrt{3}}{2}\right)} + \exp(-ik_{x}a).
    \end{split}
\end{equation}

Diagonalizing the Hamiltonian matrix $\hat{H}(\vec{k})$ gives the energy eigenvalues (band structure) $E(\vec{k}) = \pm t |S(\vec{k})| = \pm t \sqrt{3+f(\vec{k}})$, where $f(\vec{k})$ is defined below. Using an easier way $|S(\vec{k})|^2 = 3+f(\vec{k})$ and explicitly calculating $|S(\vec{k})|^2$ leads to

\begin{equation}
    \begin{split}
         |S(\vec{k})|^2 & = \left|2\exp\left(\frac{ik_{x}a}{2}\right)\cos{\left(\frac{k_{y}a\sqrt{3}}{2}\right)} + \exp(-ik_{x}a)\right|^2 \\
         & = \left[2e^{i\frac{k_{x}a}{2}} \cos\left(\frac{\sqrt{3}}{2}k_{y}a\right) + e^{-ik_{x}a} \right]\cdot\left[2e^{-i\frac{k_{x}a}{2}} \cos\left(\frac{\sqrt{3}}{2}k_{y}a\right) + e^{ik_{x}a} \right] \\
         & = 4\cos^2\left(\frac{\sqrt{3}}{2}k_{y}a\right) + 2\cos\left(\frac{\sqrt{3}}{2}k_{y}a\right)\left(e^{i\frac{3}{2}k_{x}a} + e^{-i\frac{3}{2}k_{x}a} \right) + 1 \\
         & = 4\cos^2\left(\frac{\sqrt{3}}{2}k_{y}a\right) + 4\cos\left(\frac{\sqrt{3}}{2}k_{y}a\right)\cos\left(\frac{3}{2}k_{x}a\right) + 1 \\
         & = 2 + 2\cos(2\frac{
         \sqrt{3}}{2}k_{y}a) + 4\cos\left(\frac{\sqrt{3}}{2}k_{y}a\right)\cos\left(\frac{3}{2}k_{x}a\right) + 1 \\
         & = 3 + 2\cos(\sqrt{3}k_{y}a) + 4\cos\left(\frac{\sqrt{3}}{2}k_{y}a\right)\cos\left(\frac{3}{2}k_{x}a\right).
    \end{split}
\end{equation}

Thus, the energy dispersion relation is

\begin{equation}
    E(\vec{k}) = \pm t\sqrt{3+f(\vec{k})},
\end{equation}

\noindent where

\begin{equation}
    f(\vec{k}) = 2\cos(\sqrt{3}k_y a) + 4\cos\left(\frac{\sqrt{3}}{2}k_y a\right)\cos\left(\frac{3}{2}k_x a\right).
\end{equation}

\begin{figure}[htb!]
    \centering    \includegraphics[width=0.4\textwidth]{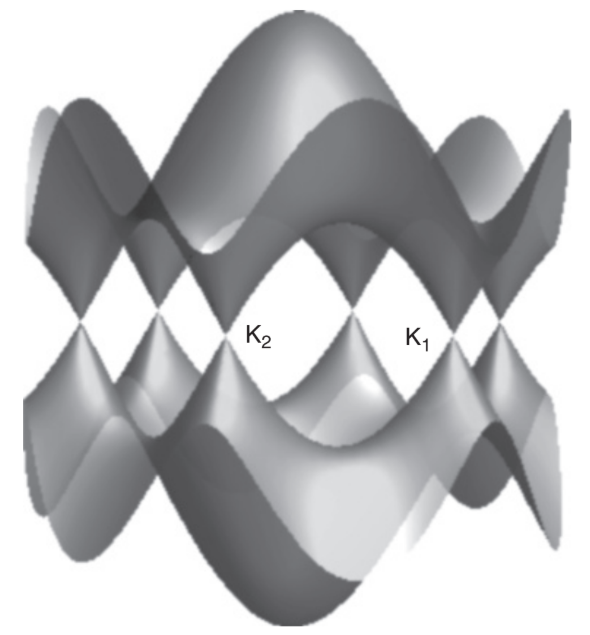}
    \caption{Electron energy spectrum of graphene in the nearest-neighbor approximation. Figure extracted from the reference \cite{graphene}.}
    \label{fig:spectrum_nn}
\end{figure}

Crucially, at the $K$ and $K'$ points defined in \autoref{eq:k-points}, the function $S(\vec{k})$ becomes zero: $S(\vec{K}) = S(\vec{K}') = 0$. This leads to $E(\vec{K}) = E(\vec{K}') = 0$, which means that the energy bands touch these specific points in the Brillouin zone. This band crossing is the origin of the unique electronic properties of graphene, as we can see in \autoref{fig:spectrum_nn}.

\section{Linear Dispersion near the Dirac Points: Emergence of Massless Dirac Fermions}

Near the $K$ and $K'$ points, the energy dispersion is linear. Expanding the Hamiltonian $\hat{H}(\vec{k})$ for small momentum deviations $\vec{q} = \vec{k} - \vec{K}$ (or $\vec{q} = \vec{k} - \vec{K}'$) leads to effective Hamiltonians

\begin{equation}
    \begin{split}
        \hat{H}_{K}(\vec{q}) &\approx \hbar v \begin{pmatrix}
        0 & q_{x}-iq_{y} \\
        q_{x}+iq_{y} & 0
        \end{pmatrix}
        = \hbar v (\sigma_x q_x + \sigma_y q_y) = \hbar v \vec{\sigma} \cdot \vec{q}, \\
        \hat{H}_{K'}(\vec{q}) &\approx \hbar v \begin{pmatrix}
        0 & q_{x}+iq_{y} \\
        q_{x}-iq_{y} & 0
        \end{pmatrix}
        = \hbar v (\sigma_x q_x - \sigma_y q_y) = \hat{H}_K^T,
    \end{split}
\end{equation}

\noindent minus a phase $\alpha =  e^{5i\pi/6}$, where $\hbar$ is the reduced Planck constant, $v = \frac{3a|t|}{2\hbar}$ is the Fermi velocity, and $\sigma_x, \sigma_y$ are Pauli matrices (note that the structures that appear naturally in these equations are Pauli structures. We emphasize that the matrices in this context arise separately from the crystal structure of the lattice and are not a consequence of imposing an SU(2) group) acting on the sublattice degree of freedom (also called pseudo-spin because it has the same matrix structure as spin, respecting the same SU(2) group theory)

\begin{equation}
    \sigma_{x} = \begin{pmatrix}
        0 & 1 \\
        1 & 0
        \end{pmatrix}, \quad
        \sigma_{y} = \begin{pmatrix}
        0 & -i \\
        i & 0
        \end{pmatrix}.
\end{equation}

We can combine the points $K$ and $K'$ in a single Hamiltonian

\begin{equation}
    \hat{H} = \begin{pmatrix}
        \hat{H}_{K} & 0 \\
        0 & \hat{H}_{K'}
    \end{pmatrix}.
\end{equation}

This Hamiltonian is mathematically identical to the Dirac Hamiltonian for massless relativistic particles in two dimensions, with the speed of light $c$ replaced by the Fermi velocity $v \approx c/300$.

The energy eigenvalues are found by diagonalizing this Hamiltonian, leading to the linear dispersion relation

\begin{equation}
    E(\vec{q}) = \pm \hbar v \sqrt{q_x^2 + q_y^2} = \pm \hbar v |\vec{q}|
\end{equation}

\begin{figure}[htb!]
     \centering     \includegraphics[width=0.6\linewidth]{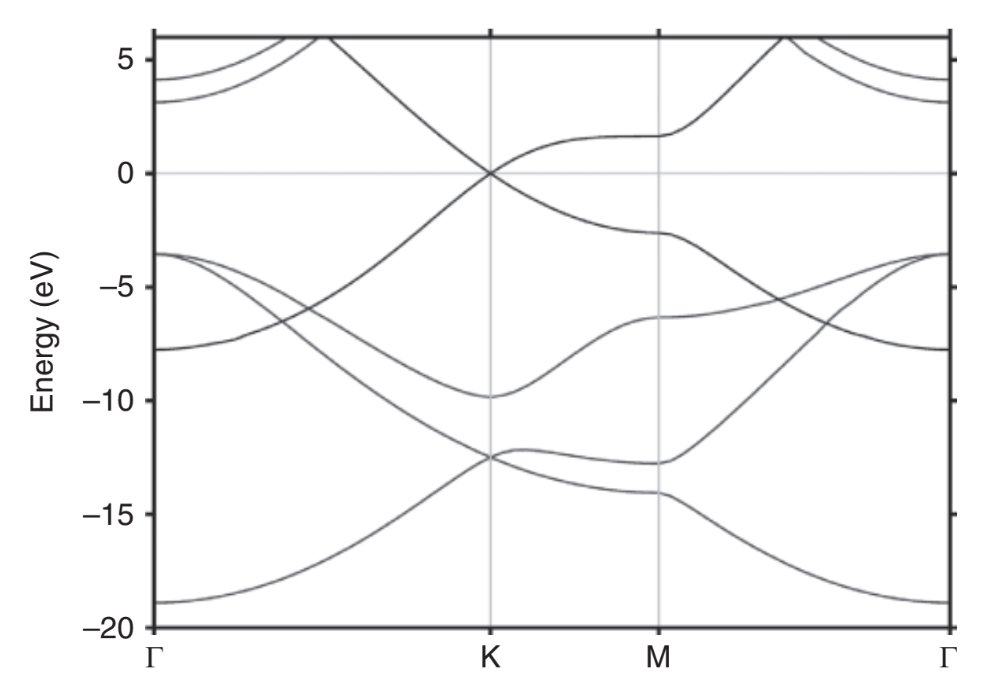}
     \caption{The band structure of graphene. Figure extracted from the reference \cite{graphene}.}
    \label{fig:dirac-cones_valence-and-conduction-bands}
 \end{figure}

This linear relationship, $E \propto |\vec{q}|$, forms the characteristic ``Dirac cones'' centered at the points $K$ and $K'$. Undoped graphene has its Fermi level precisely at these conical points ($E=0$), making it a gapless semiconductor. The Dirac cones are linear, where the conduction and valence bands touch, as we can see in the \autoref{fig:dirac-cones_valence-and-conduction-bands}.

\section{Chirality and Pseudospin}\label{subchap: chirality-and-pseudospin}








The eigenstates corresponding to the energy $E_{e,h} = \pm \hbar v |\vec{q}|$ are two-component spinors, where the components refer to the amplitudes on the A and B sublattices of the honeycomb lattice. For the $K$ valley, the electron states (e, $+$ sign) and hole (h, $-$ sign) are,

\begin{equation}
\psi_{e,h}^{(K)}(\vec{q}) = \frac{1}{\sqrt{2}} \begin{pmatrix} \exp(-i\phi_{\vec{q}}/2) \\
\pm \exp(i\phi_{\vec{q}}/2)
\end{pmatrix},
\end{equation}

\noindent and for the $K'$ valley,

\begin{equation}
\psi_{e,h}^{(K')}(\vec{q}) = \frac{1}{\sqrt{2}} \begin{pmatrix} \exp(i\phi_{\vec{q}}/2) \\
\pm \exp(-i\phi_{\vec{q}}/2)
\end{pmatrix},
\end{equation}

\noindent where $\phi_{\vec{q}}$ is the polar angle of the momentum vector $\vec{q}$. These states can now be related to chirality, a property that rigidly links the direction of their pseudo-spin to their momentum.

Recalling our discussion in chapter 2, section \ref{subchap:a-deeper-look-at-chirality-and-helicity}, where chirality was defined in the context of high-energy physics, a powerful analogue emerges in graphene. These states exhibit a remarkable property known as chirality, which rigidly links the direction of their pseudospin to their momentum. We can demonstrate this explicitly by calculating the expectation value of the pseudospin for an electron in the $K$ valley. The expectation value is given by the standard quantum mechanical formula:

\begin{equation}
\langle\vec{\sigma}\rangle=\langle\psi|\vec{\sigma}|\psi\rangle.
\end{equation}

Here, $\vec{\sigma}=(\sigma_{x},\sigma_{y})$ is the pseudospin operator, represented by the Pauli matrices which act on the sublattice space:

\begin{equation}
\sigma_{x}=\begin{pmatrix}
0 & 1 \\
1 & 0
\end{pmatrix},\quad \sigma_{y}=\begin{pmatrix}
0 & -i \\
i & 0
\end{pmatrix}.
\end{equation}

For this calculation, we can represent the electron state $\psi e^{(K)}$ by dropping a global phase factor (which does not affect the expectation value) to simplify the form:

\begin{equation}
|\psi_{+}\rangle=\frac{1}{\sqrt{2}}\begin{pmatrix}
1 \\
e^{i\phi_{\vec{q}}}
\end{pmatrix}.
\end{equation}

Calculating the expectation value for the x-component of the pseudo-spin gives:

\begin{equation}
\begin{split}
\langle\sigma_{x}\rangle =
\langle\psi_{+}|\sigma_{x}|\psi_{+}\rangle &= \frac{1}{\sqrt{2}}\begin{pmatrix}
1 & e^{-i\phi_{\vec{q}}}
\end{pmatrix}\begin{pmatrix}
0 & 1 \\
1 & 0
\end{pmatrix}\frac{1}{\sqrt{2}}\begin{pmatrix}
1 \\
e^{i\phi_{\vec{q}}}
\end{pmatrix} \\
&= \frac{1}{2}\begin{pmatrix}
e^{-i\phi_{\vec{q}}} & 1
\end{pmatrix}\begin{pmatrix}
1 \\
e^{i\phi_{\vec{q}}}
\end{pmatrix} = \frac{1}{2}\left(e^{-i\phi_{\vec{q}}}+e^{i\phi_{\vec{q}}}\right) = \cos{\phi_{\vec{q}}}.
\end{split}
\end{equation}

A similar calculation for the y-component yields:

\begin{equation}
\langle\sigma_{y}\rangle =
\langle\psi_{+}|\sigma_{y}|\psi_{+}\rangle = \sin{\phi_{\vec{q}}}.
\end{equation}

Combining these components, we find the direction of the pseudo-spin vector:

\begin{equation}
\langle\vec{\sigma}\rangle_{+} = (\cos{\phi_{\vec{q}}}, \sin{\phi_{\vec{q}}}).
\end{equation}

This result is profound: the expectation value of the pseudo-spin $\langle\sigma\rangle_{+}$ points in the exact same direction as the momentum vector $q=|q|(\cos{\phi_{q},\sin{\phi_{q}}})$. Thus, for electrons in the $K$ valley, the pseudospin is locked parallel to the momentum. An analogous calculation for electrons in the $K'$ valley shows the opposite behavior: their pseudo-spin is locked antiparallel to their momentum, confirming the opposite chirality of the two valleys.

A more formal way to express this relationship is through the chirality (helisity) operator \cite{graphene}, $(\vec{\sigma}\cdot\vec{q})/|q|$, which measures the projection of the pseudo-spin onto the momentum direction. Applying this operator to the electron and hole states confirms our finding:

\begin{equation}
    \frac{(\vec{\sigma}\cdot\vec{q})}{|\vec{q}|}\psi_{e,h} = \pm\psi_{e,h}.
\end{equation}

This signifies that for electrons (positive energy, $+$ sign), the pseudo-spin is parallel to momentum, while for holes (negative energy, $-$ sign), it is antiparallel. This fixed relationship defines the chirality of these massless Dirac fermions.

\section{Beyond the Simplest Model: Next-Nearest Neighbors and Trigonal Warping}

Including hopping between next-nearest neighbors ($t'$) modifies the tight-binding energy to 

\begin{equation}
    E(\vec{k}) = \pm t\sqrt{3+f(\vec{k})} + t'f(\vec{k}).
\end{equation}

This term $t'$ breaks the perfect electron-hole symmetry, slightly shifting the Dirac point energy to $E = -3t'$, but it does not open an energy gap. Given typical values ($t \approx -2.97\ eV$, $t' \approx -0.073\ eV$), $t'$ is small, which makes the electron-hole symmetry a good approximation.

Higher-order terms in the expansion near $K$ and $K'$ introduce effects such as the electron energy bands becoming triangular (trigonal warping). The Hamiltonian including the next quadratic term can be written as

\begin{equation}
    \hat{H} = \hbar v \tau_{0}\otimes\vec{\sigma}\cdot\vec{k} + \mu\tau_{z}\otimes[2\sigma_{y}k_{x}k_{y}-\sigma_{x}({k_{x}}^{2}-{k_{y}}^{2})],
\end{equation}

\noindent where $\tau_0, \tau_z$ act on the valley index and $\mu=3a^{2}t/8$. This additional term makes the energy dispersion anisotropic 

\begin{equation}
    \epsilon^{2}(\vec{k}) = \hbar^{2}v^{2}k^{2} \mp 2\hbar v\mu k^{3}\cos(3\phi_{\vec{k}}) + \mu^{2}k^{4},
\end{equation}

\noindent introducing a threefold symmetry (warping) to the Dirac cones.

\section{Symmetry Protection of the Gapless State}

The gapless nature of graphene at the Dirac points is protected by time-reversal ($T$) and inversion ($I$) symmetry. These symmetries impose conditions relating the Hamiltonians at $K$ and $K'$ and lead to a constraint on the Hamiltonian within a single valley

\begin{equation}
    TI \implies H_{K} = \sigma_{x}H_{K}^{*}\sigma_{x}
\end{equation}

Writing $H_K = \sum_{i=0,x,y,z}\alpha_{i}\sigma_{i}$, this constraint forces the coefficient of $\sigma_z$ to be zero ($\alpha_z = 0$). Since the $\sigma_z$ term corresponds to a mass term that would open a gap, its absence ensures that the gapless state is protected as long as the $T$ and $I$ symmetries hold. Breaking inversion symmetry (e.g., via a substrate) allows a mass term and can open a gap.

\section{Symmetries in Graphene}

\subsection{Crystal Symmetry in Graphene}

The graphene lattice can be described by its crystal symmetry, a two-dimensional symmetry group that reflects the periodicity of the honeycomb structure. The hexagonal symmetry of graphene leads to the presence of two non-equivalent sublattices, often labeled $A$ and $B$. These sublattices are related to each other by a mirror symmetry and are the key to understanding many of the electronic properties of the material. The symmetry of the lattice underpins the behavior of electrons in graphene and influences the electronic band structure, which features a characteristic linear energy dispersion near the Dirac points ($K$ and $K'$ points in the Brillouin zone).

\subsection{Sublattice Symmetry and Dirac Fermions}

The electronic structure of graphene is deeply influenced by its sublattice symmetry. The Dirac-like behavior of electrons in graphene arises from the symmetry between the $A$ and $B$ sublattices. This symmetry implies that the low-energy excitations in graphene behave like massless Dirac fermions, leading to a linear dispersion relation in the vicinity of the Dirac points. The symmetry between the two sublattices ensures that, at low energies, the conduction and valence bands touch at these Dirac points, forming a point of zero-energy gap. This results in characteristic properties of graphene, such as high electron mobility and a high degree of conductivity.

In addition to the sublattice symmetry, the crystal symmetry of graphene also dictates the specific form of the Hamiltonian that describes the behavior of the electrons. The matrix elements of the Hamiltonian in momentum space reflect the symmetry of the lattice and are responsible for the massless Dirac spectrum observed in graphene.

\subsection{Time-Reversal Symmetry}

The time-reversal symmetry ($T$) is another fundamental symmetry present in graphene. This symmetry, which inverts the direction of time, plays an important role in the electronic properties of graphene, particularly in the absence of magnetic fields. When graphene is not subject to any external perturbation such as a magnetic field, time-reversal symmetry ensures that the material exhibits time-reversal invariance, meaning that the evolution of the system is unchanged when time is reversed. This symmetry is crucial for the protection of certain topological states, such as the quantum spin Hall effect, which could be realized in graphene-based materials with spin-orbit coupling.

\subsection{Symmetry Breaking and Topological States}

Symmetry breaking in graphene can lead to the emergence of novel physical phenomena. One of the most notable examples is the possible breaking of the sublattice symmetry, which could result in the opening of an energy gap at the Dirac points. This can occur under certain conditions, such as the application of a strong external electric field, leading to the formation of a non-trivial topologically insulating state known as a topological insulator. Additionally, the introduction of a staggered sublattice potential or a magnetic field can break the time-reversal symmetry, leading to the formation of edge states that are protected by topological invariants.

Another form of symmetry breaking occurs because of the interaction effects in graphene. When electron-electron interactions become significant, the material may undergo a phase transition, breaking the crystal symmetry and leading to the formation of new ordered phases, such as a charge density wave or a superconducting phase. These symmetry-breaking phenomena are crucial for understanding the rich phase diagram of graphene and graphene-based materials.

In the next chapter, having established that the inherent symmetries of pristine graphene protect its gapless state and thus present a fundamental obstacle, we will construct a theoretical model to overcome this limitation. We will move beyond pristine graphene by proposing a system based on an engineered honeycomb lattice where sublattice symmetry is explicitly broken. This crucial modification introduces the mass gap required to define robust pseudo-chiral states , thereby laying the necessary groundwork to investigate the conditions under which a $2+1D$ analogue of the CME can emerge.

\chapter{Toward the CME - A Model for Valley Imbalance}
The CME is a remarkable quantum phenomenon predicted to occur in $3+1D$ systems hosting relativistic fermions, such as the QGP or Weyl semimetals. At its core, the effect is deeply connected to the concept of chirality, a quantum property of fermions defined by the $\gamma^{5}$ matrix, which acts as the chiral operator. The quantum state of a particle, its spinor, can be an eigenvector of $\gamma^{5}$ with an eigenvalue of $+1$ (defining a right-handed state) or $-1$ (a left-handed state). The $\gamma^{5}$ matrix, defined in four-dimensional spacetime as $\gamma^{5} = i \gamma^{0}\gamma^{1}\gamma^{2}\gamma^{3}$, possesses a crucial property: it must anticommute with the other gamma matrices ($\{\gamma^{5},\gamma^{\mu} \}=0$). This ensures that the spinor space is cleanly divided into two distinct subspaces for left- and right-handed particles. The CME then manifests as the generation of an electric current parallel to an applied magnetic field, arising from a non-equilibrium imbalance between these left- and right-handed chiral populations.

A direct translation of this phenomenon to lower dimensions, however, confronts a fundamental obstacle. The chiral operator $\gamma^5$ is inherently a four-dimensional construct, as its definition requires the product of four gamma matrices. In a $2+1D$ system, which only has three gamma matrices, it is impossible to construct an equivalent operator that anticommutes with all of them. The central motivation of this thesis is therefore to investigate whether an analogous effect, built upon a different definition of ``pseudo-chirality'', can be engineered and observed in a $2+1D$ material platform.

The natural starting point for such an investigation is pristine graphene, a material celebrated for its low-energy electronic structure described by two cones of massless Dirac fermions. However, inherent high symmetry graphene presents a fundamental obstacle. Specifically, the inversion symmetry between its two carbon sublattices ($A$ and $B$) ensures they are physically and energetically identical. This symmetry protects the gapless nature of the Dirac points and prevents the formation of a mass gap, which is a prerequisite for defining the distinct ``chiral-like'' states needed to replicate the CME.

\begin{figure}[htb!]
     \centering
     \includegraphics[width=0.4\linewidth]{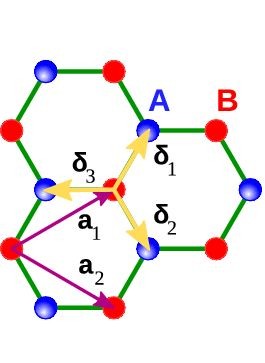}
     \caption{Honeycomb lattices where the two sublattice sites are occupied by different atomic species}
     \label{fig:lattice-graphene}
 \end{figure}

To construct a viable $2D$ analogue of the CME, we must therefore break this sublattice symmetry. We turn our attention to honeycomb lattices where the two sublattice sites are occupied by different atomic species, as depicted in \autoref{fig:lattice-graphene}, such as in monolayer boron nitride (BN)\footnote{The boron nitride (BN) itself is an insulator, because of that it is used a mix of graphene and BN to create a material with the properties desired to our investigations.}. In this structure, the Boron and Nitrogen atoms create an intrinsic on-site energy potential difference, explicitly breaking the inversion symmetry that constrained pristine graphene.

\begin{figure}[htb!]
\centering
\includegraphics[width=0.4\linewidth]{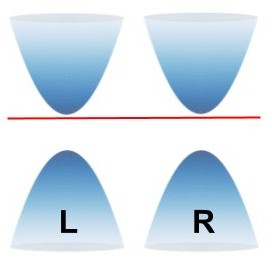}
\caption{Gapped system}
\label{fig:gapped-system}
\end{figure}

The low-energy charge carriers in one of the valleys of such a gapped system [\autoref{fig:gapped-system}] can be effectively described by the massive $2D$ Dirac Hamiltonian,

\begin{equation}
\mathcal{H} = \hbar v_F (k_x \sigma_x + k_y \sigma_y) + m \sigma_z.
\end{equation}

In this $2D$ framework, as mentioned in section \ref{subchap: chirality-and-pseudospin}, the cones $K$ and $K'$ assume a role directly analogous to the $\gamma^{5}$ matrix in $3+1D$ systems. The mass term, $\mathcal{H}_{m}=m\sigma_{z}$, is what opens a band gap of magnitude $2|m|$, allowing the valley index to be defined as a new, pseudo-chiral quantum number. While the operator $\sigma_z$ in the mass term leads to states with a well-defined sublattice polarization — a higher probability of residing on one sublattice over the other — it is the valley index itself ($K$ and $K'$) that serves as the direct and powerful analog to the left- and right-handedness of relativistic fermions.

We propose to engineer a non-equilibrium imbalance in the pseudo-chiral populations (i.e., an excess of carriers localized in $K$ and $K'$). For the first step, we will add a $\Delta$ with exchanged signs, using the reference from \cite{PhysRevLett.106.116803}, and investigate whether this engineered imbalance, when coupled with an external magnetic field, can generate a net electrical current parallel to the field, thereby realizing a genuine $2+1D$ CME.

Starting with the equation \textbf{6b} from \cite{PhysRevLett.106.116803}, in which we have a Hamiltonian \footnote{This Hamiltonian covers the lattice and the sub lattice in its terms, knowing that the honeycomb structure is considered in two lattices that work together, showing specific properties of electrons displacement.} that describes how the electrons behave in this honeycomb lattice in a specific orientation \(\mathcal{P}\),

\begin{equation}
    \mathcal{H}_{P} = v_{F}(\kappa\sigma_{x}p_{x}+\sigma_{y}p_{y})+\sigma_{z}\Delta,
    \label{papper_equation_6b}
\end{equation}

\noindent where \(v_{F}\) is the Fermi velocity, \(\kappa\) is the ``valley'' index that distinguishes the two types of non-equivalent \(\mathbf{K}^{\pm}\) points, \(\sigma_{j}\) are the Pauli matrices and \(p_{j}\) is the momentum of the particles in their respective directions. These \(\mathbf{K}\) points are the points where the Dirac cones touch itself, considered by the energy band structure point of view. 

The first two terms of this equation are well known and follow the prescription used in \cite{graphene}, in which the properties of the honeycomb structure of graphene are described. Presently, we are interested in studying this Hamiltonian, but considering a z-projection, which a priori is not considered in graphene. This z-projection exists in other Dirac and Weyl materials, which are of our interest. This is represented in the last term of the Hamiltonian in \autoref{papper_equation_6b}. The \(\Delta\) parameter represents a gap opening between the valence and conduction bands. It appears in both \(\mathbf{K}\) and \(\mathbf{K'}\) indicating an identical gap for both pairs of cones.

\section{Model for Valley Imbalance}

\begin{figure}[htb!]
    \centering
    \includegraphics[width=0.4\linewidth]{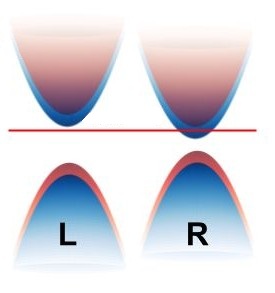}
    \caption{Imbalance cone}
    \label{fig:imbalance-cone}
\end{figure}

Considering these two pairs, we can make an analogous labeling one of the pairs as $L$ and the other as $R$. To simulate the chiral imbalance that exists in the CME — where there is an excess of left- over right-handed quarks, or vice-versa — we aim to create a corresponding imbalance between these pairs of cones. This is achieved by introducing a new parameter, $\delta$, which represents a break in time-reversal symmetry, that sums with the pre-existing gap parameter, $\Delta$. By applying this $\delta$ term with opposite signs to each cone, one gap is made wider while the other becomes narrower \cite{Mizher:2018dtf,Dudal:2021ret} (see \autoref{fig:imbalance-cone}). This method establishes the non-equilibrium condition between the valleys that is central to our investigation

Before proposing the addition of the parameter \(\delta\), it will be interesting to our work to explore a concept proposed in \cite{PhysRevLett.106.116803}. We already start with \autoref{papper_equation_6b}, and this will be called the ``lattice spin''.

\subsection{THE Lattice Spin ($\mathbf{S}$)}

The concept of lattice spin ($\mathbf{S}$) is a key element in the study of Dirac and Weyl materials. It represents a unique form of angular momentum that emerges from the orbital motion of the electrons as it interacts with the specific geometry of the honeycomb lattice. Rather than being a property of the electron itself, it is a characteristic of the quasi-particle — the electron coupled to its lattice environment.  This makes the operator $\mathbf{S}$ fundamentally different from both the conventional electron spin, which is an intrinsic property of the particle, and the dimensionless pseudo-spin, which is a mathematical label for the sublattice degree of freedom (i.e., whether the electron is on sublattice $A$ or $B$).

This distinction is crucial because it explains why electrons in these materials behave like relativistic-like fermions. The lattice spin is the missing piece that ensures the total angular momentum of the quasi-particle ($\mathbf{L} + \mathbf{S}$) is a conserved quantity, even when its orbital angular momentum ($\mathbf{L}$) alone is not.

\begin{figure}[htb!]
     \centering
     \includegraphics[width=0.4\linewidth]{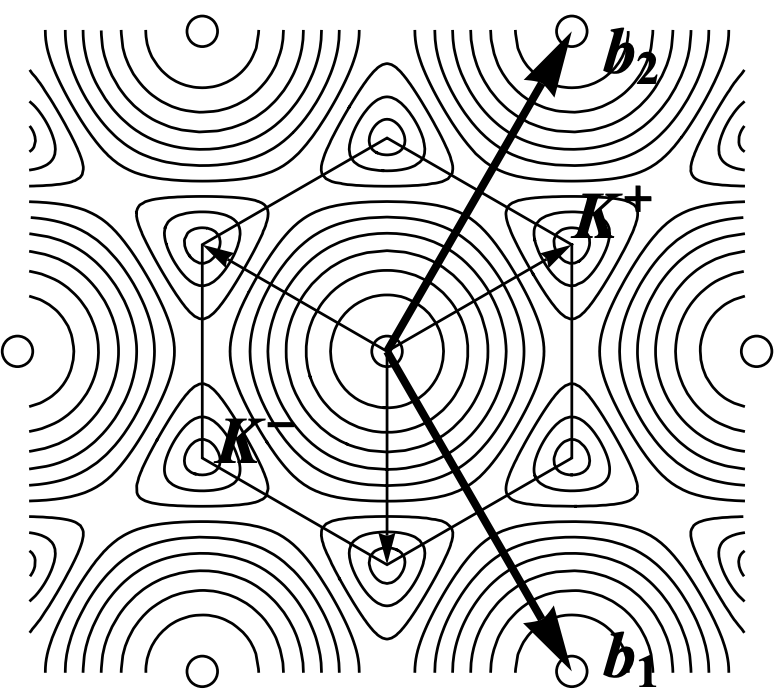}
     \caption{Honeycomb lattice band structure}
     \label{fig:brillouin}
 \end{figure}

Observing the first Brilloin zone, where the $K$ and $K'$ points are located, we notice that around these points there is a rotational symmetry, represented in \autoref{fig:brillouin}. This in principle indicated a conserved quantity, which is not the case, as we will show below. It becomes conservative only with the addition of the lattice spin. The sum of the angular momentum and the lattice spin, commutes with the Hamiltonian of the honeycomb lattice, and we still have this rotational symmetry conserved.

\subsection{Understanding the Hamiltonian Behavior with $\mathbf{S}$}

Now we are going to reproduce this calculation, then we add the proposed time-reversal symmetry break with the parameter $\delta$, to finally couple a magnetic field on our Hamiltonian to analyze the potential occurrence of the CME in honeycomb lattices with broken sublattice symmetry.

A key principle in quantum mechanics connects the symmetries of a system to its conserved quantities. For the rotational symmetry observed in the Brillouin zone [\autoref{fig:brillouin}], the associated conserved quantity is the angular momentum ($\hat{L}$), which is the generator of rotations. An operator is conserved if its time evolution is zero, and the Heisenberg equation of motion shows that this time evolution is governed by the commutator of the operator with the Hamiltonian ($\hat{H}$):

\begin{equation}
    \frac{d\hat{L}}{dt} = \frac{i}{\hbar}[\hat{H}, \hat{L}]
\end{equation}

Therefore, for angular momentum to be conserved ($d\hat{L}/dt = 0$), its commutator with the Hamiltonian must vanish. Verifying if the rotational symmetry holds in our model translates directly into testing this mathematical condition.

We begin by performing this check, calculating the commutator between the Hamiltonian from \autoref{papper_equation_6b} and the angular momentum operator. We expect this to be a conserved quantity, and we can determine the conditions for this by seeing when the commutator vanishes:

\begin{equation}
    [\mathcal{H}_{P},\vec{\mathbf{L}}]
\end{equation}

Using these commutator calculations split by coordinates, we have:

\begin{equation}
    \begin{split}
        [\mathcal{H}_{P}, \mathbf{L}_{x}] & = [(v_{F}\kappa\sigma_{x}p_{x} + v_{F}\sigma_{y}p_{y} + \sigma_{z}\Delta) \cdot (r_{y}p_{z}-r_{z}p_{y})] \\
        & - [(r_{y}p_{z}-r_{z}p_{y}) \cdot (v_{F}\kappa\sigma_{x}p_{x} + v_{F}\sigma_{y}p_{y} + \sigma_{z}\Delta)] \\ 
        & = (v_{F}\kappa\sigma_{x}p_{x}r_{y}p_{z}-v_{F}\kappa\sigma_{x}p_{x}r_{z}p_{y}+v_{F}\sigma_{y}p_{y}r_{y}p_{z}-v_{F}\sigma_{y}p_{y}p_{z}p_{y} + \sigma_{z}\Delta r_{y}p_{z}-\sigma_{z}\Delta r_{z}p_{y}) \\ 
        & - (r_{y}p_{z}v_{F}\kappa\sigma_{x}p_{x}+r_{y}p_{z}v_{F}\sigma_{y}p_{y}+r_{y}p_{z}\sigma_{z}\Delta -r_{z}p_{y}v_{F}\kappa\sigma_{x}p_{x}-r_{z}p_{y}v_{F}\sigma_{y}p_{y}-r_{z}p_{y}\sigma_{z}\Delta) \\
        & = i\hbar v_{F}\sigma_{y}p_{z},
    \end{split}
\end{equation}

\begin{equation}
    \begin{split}
        [\mathcal{H}_{P}, \mathbf{L}_{y}] & = [(v_{F}\kappa\sigma_{x}p_{x} + v_{F}\sigma_{y}p_{y} + \sigma_{z}\Delta) \cdot (-r_{x}p_{z} + r_{z}p_{y})] \\
        & - [(-r_{x}p_{z} + r_{z}p_{y}) \cdot (v_{F}\kappa\sigma_{x}p_{x} + v_{F}\sigma_{y}p_{y} + \sigma_{z}\Delta)] \\
        & = (-v_{F}\kappa\sigma_{x}p_{x}r_{x}p_{z} + v_{F}\kappa\sigma_{x}p_{x}r_{z}p_{y} - v_{F}\sigma_{y}p_{y}r_{x}p_{z} + v_{F}\sigma_{y}p_{y}r_{z}p_{y} - \sigma_{z}\Delta r_{x}p_{z} + \sigma_{z}\Delta r_{z}p_{y}) \\
        & - (-r_{x}p_{z}v_{F}\kappa\sigma_{x}p_{x} - r_{x}p_{z}v_{F}\sigma_{y}p_{y} - r_{x}p_{z}\sigma_{z}\Delta + r_{z}p_{y}v_{F}\kappa\sigma_{x}p_{x} + r_{z}p_{y}v_{F}\sigma_{y}p_{y} + r_{z}p_{y}\sigma_{z}\Delta) \\
        & = i\hbar v_{F}\kappa\sigma_{x}p_{z},
    \end{split}
\end{equation}

\begin{equation} \label{papper_hamiltonian_angular_momentum_z}
    \begin{split}
        [\mathcal{H}_{P},\mathbf{L}_{z}] & = [(v_{F}\kappa\sigma_{x}p_{x} + v_{F}\sigma_{y}p_{y} + \sigma_{z}\Delta) \cdot (r_{x}p_{y} - r_{y}p_{x})] \\
        & - [(r_{x}p_{y} - r_{y}p_{x}) \cdot (v_{F}\kappa\sigma_{x}p_{x} + v_{F}\sigma_{y}p_{y} + \sigma_{z}\Delta)] \\\
        & = (v_{F}\kappa\sigma_{x}p_{x}r_{x}p_{y} - v_{F}\kappa\sigma_{x}p_{x}r_{y}p_{x} + v_{F}\sigma_{y}p_{y}r_{x}p_{y} - v_{F}\sigma_{y}p_{y}r_{y}p_{x} + \sigma_{z}\Delta r_{x}p_{y} - \sigma_{z}\Delta r_{y}p_{x}) \\
        & - (r_{x}p_{y}v_{F}\kappa\sigma_{x}p_{x} + r_{x}p_{y}v_{F}\sigma_{y}p_{y} + r_{x}p_{y}\sigma_{z}\Delta - r_{y}p_{x}v_{F}\kappa\sigma_{x}p_{x} - r_{y}p_{x}v_{F}\sigma_{y}p_{y} - r_{y}p_{x}\sigma_{z}\Delta) \\
        & = v_{F}\kappa\sigma_{x}(-i\hbar p_{y}) + v_{F}\sigma_{y}(i\hbar p_{x}),
    \end{split}
\end{equation}

\noindent in a compact way,

\begin{equation}
    [\mathcal{H}_{p},\vec{\mathbf{L}}] = -i\hbar v_{F}\begin{pmatrix}
        \sigma_{y}p_{z} \\
        -\kappa\sigma_{x}p_{z} \\
        \kappa\sigma_{x}p_{y}-\sigma_{y}p_{x}
    \end{pmatrix}.
\end{equation}

The fact that the z-component of the commutator does not vanish despite the rotational symmetry around this axis suggests that some components of angular momentum in this direction is missing. The reference \cite{PhysRevLett.106.116803} therefore suggest that a new quantity, the lattice spin, can be defined in such a way that it complements the angular momentum in the z-direction, allowing the commutator of the total angular momentum and the Hamiltonian to vanish.

After proving the results of the commutators between the Hamiltonian and the angular momentum, we start the calculus of the commutators between the Hamiltonian and the \(\mathbf{S}\) operator, and the coordinate independent notation known as \(\mathbf{u}\), as described below:

\begin{equation} \label{papper_lattice_spin}
    \mathbf{S} \equiv  \frac{\hbar}{2} (\kappa\sigma_{x}\hat{\mathbf{a}}_{d} + \sigma_{y}\hat{\mathbf{a}}_{s} + \kappa\sigma_{z}\hat{\mathbf{a}}_{n}),
\end{equation}

\begin{equation} \label{papper_coodinates_independent}
    \mathbf{u} \equiv (\mathbf{p}\cdot\hat{\mathbf{a}}_{d})\hat{\mathbf{a}}_{d} + (\mathbf{p}\cdot\hat{\mathbf{a}}_s)\hat{\mathbf{a}}_{s} + \kappa(\Delta/v_{F})\hat{\mathbf{a}}_{n}.
\end{equation}

In coordinate-independent notation, the honeycomb Hamiltonian is:

\begin{equation} \label{papper_coordinate-independent_notation_the_honeycomb_hamiltonian}
    \mathcal{H} = \frac{2v_{F}}{\hbar}\mathbf{S}\cdot\mathbf{u}.
\end{equation}

By calculating the scalar product of \(\mathbf{S}\) and \(\mathbf{u}\),

\begin{equation} \label{papper_scalar_product_S_u}
    \mathbf{S}\cdot\mathbf{u} = \frac{\hbar}{2} [\kappa\sigma_{x}p_{x} + \sigma_{y}p_{y} + \sigma_{z}\Delta/v_{F}],
\end{equation}

\noindent and replacing in \autoref{papper_coordinate-independent_notation_the_honeycomb_hamiltonian} we obtain, 

\begin{equation}
    \mathcal{H} = \frac{2v_{F}}{\hbar}\frac{\hbar}{2} [\kappa\sigma_{x}p_{x} + \sigma_{y}p_{y} + \sigma_{z}\Delta/v_{F}]=v_{F}\kappa\sigma_{x}p_{x}+v_{F}\sigma_{y}p_{y}+\sigma_{z}\Delta.
\end{equation}

Continuing our calculations, we can verify if the affirmation,

\begin{equation} \label{papper_commutator_hamiltonian_angular-momentum_s}
    [\mathcal{H}, (\mathbf{L} + \mathbf{S})\cdot \hat{\mathbf{a}}_{n}] = 0
\end{equation}

\noindent for any value of \(\Delta\), is true.

To prove the assertion in \autoref{papper_commutator_hamiltonian_angular-momentum_s}, we need to evaluate the commutator of the Hamiltonian with the z-component of the total angular momentum, $(\mathbf{L} + \mathbf{S})_z$. We will begin by calculating the commutator with the lattice spin component, $[\mathcal{H}, \mathbf{S}_{z}]$.

So let us first perform the Hamiltonian commutation with the \(\mathbf{S}\) operator in the z coordinate:

\begin{equation} \label{papper_hamiltonian_s}
    \begin{split}
        [\mathcal{H}, \mathbf{S}_{z}] & = [(v_{F}\kappa\sigma_{x}p_{x} + v_{F}\sigma_{y}p_{y} + \sigma_{z}\Delta) \cdot (\hbar/2\kappa\sigma_{z})] \\
        & - [(\hbar/2\kappa\sigma_{z}) \cdot (v_{F}\kappa\sigma_{x}p_{x} + v_{F}\sigma_{y}p_{y} + \sigma_{z}\Delta)] \\
        & = (v_{F}\kappa\sigma_{x}p_{x}\hbar/2\kappa\sigma_{z} + v_{F}\sigma_{y}p_{y}\hbar/2\kappa\sigma_{z} + \sigma_{z}\Delta\hbar/2\kappa\sigma_{z}) \\
        & - (\hbar/2\kappa\sigma_{z}v_{F}\kappa\sigma_{x}p_{x} - \hbar/2\kappa\sigma_{z}v_{F}\sigma_{y}p_{y} - \hbar/2\kappa\sigma_{z}\sigma_{z}\Delta) \\
        & = v_{F}\kappa\sigma_{x}p_{x}\hbar/2\kappa\sigma_{z} + v_{F}\sigma_{y}p_{y}\hbar/2\kappa\sigma_{z} - \hbar/2\kappa\sigma_{z}v_{F}\kappa\sigma_{x}p_{x} - \hbar/2\kappa\sigma_{z}v_{F}\sigma_{y}p_{y} \\
        & = v_{F}p_{x}\kappa^2i\hbar\sigma_{y} - v_{F}p_{y}\kappa i\hbar\sigma_{x}.
    \end{split}
\end{equation}

We can now use \autoref{papper_hamiltonian_angular_momentum_z} and \autoref{papper_hamiltonian_s}, both on the z-axis coordinates, to verify the statement in \autoref{papper_commutator_hamiltonian_angular-momentum_s}.

Performing the calculus of the commutator \eqref{papper_commutator_hamiltonian_angular-momentum_s},

\begin{equation}
    \begin{split}
        [\mathcal{H}, (\mathbf{L}_{z} + \mathbf{S}_{z})] & = [\mathcal{H}, \mathbf{L}_{z}] + [\mathcal{H}, \mathbf{S}_{z}] \\
        & = (-v_{F}\kappa\sigma_{x}i\hbar p_{y} + v_{F}\sigma_{y}i\hbar p_{x}) \\
        & + (v_{F}p_{x}\kappa^2i\hbar\sigma_{y} - v_{F}p_{y}\kappa i\hbar\sigma_{x}) \\
        & = i\hbar v_{F}\kappa^2\sigma_{y}p_{x}-i\hbar v_{F}\sigma_{y}p_{x} \\
        & = i\hbar v_{F}\sigma_{y}p_{x}(\kappa^2-1),
    \end{split}
\end{equation}

\noindent and knowing that \(\kappa = \pm 1\) can have only these two values, the result will be,  

\begin{equation}
    [\mathcal{H}, (\mathbf{L}_{z} + \mathbf{S}_{z})] = 0,
    \label{eq:conserved-ang-mom}
\end{equation}

\noindent as predicted in the paper\cite{PhysRevLett.106.116803}.

As we have analyzed, by breaking down all the equations, we can see that the statement \([\mathcal{H}, (\mathbf{L}_{z} + \mathbf{S}_{z})] = 0\) for any value of \(\Delta\) is true.


\section{Engineering the Valley Imbalance}

Having validated the physical consistency of the gapped Hamiltonian by confirming the conservation of total angular momentum [\autoref{eq:conserved-ang-mom}], the next essential step is to introduce the primary ingredient for a CME analogue: a non-equilibrium imbalance between the valley populations.

To achieve this, our strategy is to asymmetrically modify the energy gaps of the two Dirac cones. This is accomplished by introducing a new parameter $\delta$, which is not merely a mathematical tool but has a profound physical meaning as a term that explicitly breaks time-reversal symmetry.

By adding this $\delta$ parameter to the existing gap $\Delta$ with opposite signs for the $K^{+}$ and $K^{-}$ valleys, we achieve the desired deformation. This procedure widens the gap in one valley while narrowing it in the other, creating a controllable imbalance known as valley polarization.

Our proposal is to add this gap as follow,

\begin{equation}
    \mathbf{u'} \equiv (\mathbf{p}\cdot\hat{\mathbf{a}}_{d})\hat{\mathbf{a}}_{d} + (\mathbf{p}\cdot\hat{\mathbf{a}}_s)\hat{\mathbf{a}}_{s} + \kappa(\Delta/v_{F} + \delta\kappa)\hat{\mathbf{a}}_{n},
\end{equation}

\noindent where we add another energy difference to the already existing \(\Delta\), forcing an even larger and asymmetric opening between the pair of cones.

Having understood our first proposal, we need to redo the commutator calculations for the new Hamiltonian that will emerge incorporating the \(\delta\) parameter in it, this will be done redefining the independent coordinates \(\mathbf{u}\). Next, we will redo these calculations.

Calculating the scalar product of \(\mathbf{S}\) and \(\mathbf{u'}\),  

\begin{equation}
    \mathbf{S}\cdot\mathbf{u'} = \frac{\hbar}{2} [\kappa\sigma_{x}p_{x} + \sigma_{y}p_{y} + \sigma_{z}(\Delta/v_{F} + \delta\kappa)].
\end{equation}

This expression, if we combine the pairs $K$ and $K'$ in the same matrix, can be represented by,

\begin{equation}
     \scalebox{0.8}{$\mathcal{H}_{\mathbf{K, K'}} = \left(\begin{matrix}
        \Delta + \delta & \sqrt{3}ta\mathbf{k}(\kappa\hat{\mathbf{a}}_{d}-i\hat{\mathbf{a}}_{s}/2) & 0 & 0 \\
        \sqrt{3}ta\mathbf{k}(\kappa\hat{\mathbf{a}}_{d}+i\hat{\mathbf{a}}_{s}/2) & -\Delta - \delta & 0 & 0 \\
        0 & 0 & \Delta - \delta & \sqrt{3}ta\mathbf{k}(\kappa\hat{\mathbf{a}}_{d}+i\hat{\mathbf{a}}_{s}/2) \\
        0 & 0 & \sqrt{3}ta\mathbf{k}(\kappa\hat{\mathbf{a}}_{d}-i\hat{\mathbf{a}}_{s}/2) & -\Delta + \delta
    \end{matrix}\right)$}.
    \label{eq:hamiltonian-pairs-combined}
\end{equation}

The honeycomb Hamiltonian will be,

\begin{equation}
    \begin{split}
        \mathcal{H'} & = \frac{2v_{F}}{\hbar}\frac{\hbar}{2} [\kappa\sigma_{x}p_{x} + \sigma_{y}p_{y} + \sigma_{z}(\Delta/v_{F} + \delta\kappa)] \\
        & = v_{F}\kappa\sigma_{x}p_{x} + v_{F}\sigma_{y}p_{y} + \sigma_{z}\Delta + v_{F}\sigma_{z}\delta\kappa.
    \end{split}
\end{equation}

Now we want to verify whether this manipulation \cite{Li2013} is valid by calculating the commutator of \(\mathcal{H'}\) and \(\mathbf{L}_{z}\),

\begin{equation}
    \begin{split}
        [\mathcal{H'}, \mathbf{L}_{z}] & = [(v_{F}\kappa\sigma_{x}p_{x} + v_{F}\sigma_{y}p_{y} + \sigma_{z}\Delta + v_{F}\sigma_{z}\delta\kappa) \cdot (r_{x}p_{y} - r_{y}p_{x})] \\
        & - [(r_{x}p_{y} - r_{y}p_{x}) \cdot (v_{F}\kappa\sigma_{x}p_{x} + v_{F}\sigma_{y}p_{y} + \sigma_{z}\Delta + v_{F}\sigma_{z}\delta\kappa)] \\
        & = (v_{F}\kappa\sigma_{x}p_{x}r_{x}p_{y} - v_{F}\kappa\sigma_{x}p_{x}r_{y}p_{x} + v_{F}\sigma_{y}p_{y}r_{x}p_{y} - v_{F}\sigma_{y}p_{y}r_{y}p_{x} \\ 
        & + \sigma_{z}\Delta r_{x}p_{y} - \sigma_{z}\Delta r_{y}p_{x} + v_{F}\sigma_{z}\delta\kappa r_{x}p_{y} - v_{F}\sigma_{z}\delta\kappa r_{y}p_{x}) \\ 
        & - (r_{x}p_{y}v_{F}\kappa\sigma_{x}p_{x} - r_{y}p_{x}v_{F}\kappa\sigma_{x}p_{x} + r_{x}p_{y}v_{F}\sigma_{y}p_{y} - r_{y}p_{x}v_{F}\sigma_{y}p_{y} \\
        & + r_{x}p_{y}\sigma_{z}\Delta - r_{y}p_{x}\sigma_{z}\Delta + r_{x}p_{y}v_{F}\sigma_{z}\delta\kappa - r_{y}p_{x}v_{F}\sigma_{z}\delta\kappa) \\
        & = v_{F}\kappa\sigma_{x}(-i\hbar p_{y}) + v_{F}\sigma_{y}(i\hbar p_{x}),
    \end{split}
\end{equation}

\noindent and the commutator of \(\mathcal{H'}\) with \(\mathbf{S}\),

\begin{equation}
    \begin{split}
        [\mathcal{H'}, \mathbf{S}_{z}] & = [(v_{F}\kappa\sigma_{x}p_{x} + v_{F}\sigma_{y}p_{y} + \sigma_{z}\Delta + v_{F}\sigma_{z}\delta\kappa) \cdot (\hbar/2\kappa\sigma_{z})] \\
        & - [(\hbar/2\kappa\sigma_{z}) \cdot (v_{F}\kappa\sigma_{x}p_{x} + v_{F}\sigma_{y}p_{y} + \sigma_{z}\Delta + v_{F}\sigma_{z}\delta\kappa)] \\
        & = (v_{F}\kappa\sigma_{x}p_{x}\hbar/2\kappa\sigma_{z} + v_{F}\sigma_{y}p_{y}\hbar/2\kappa\sigma_{z} + \sigma_{z}\Delta\hbar/2\kappa\sigma_{z} + v_{F}\sigma_{z}\delta\kappa\hbar/2\kappa\sigma_{z}) \\
        & - (\hbar/2\kappa\sigma_{z}v_{F}\kappa\sigma_{x}p_{x} - \hbar/2\kappa\sigma_{z}v_{F}\sigma_{y}p_{y} - \hbar/2\kappa\sigma_{z}\sigma_{z}\Delta + \hbar/2\kappa\sigma_{z}v_{F}\sigma_{z}\delta\kappa) \\
        & = v_{F}\kappa\sigma_{x}p_{x}\hbar/2\kappa\sigma_{z} + v_{F}\sigma_{y}p_{y}\hbar/2\kappa\sigma_{z} - \hbar/2\kappa\sigma_{z}v_{F}\kappa\sigma_{x}p_{x} - \hbar/2\kappa\sigma_{z}v_{F}\sigma_{y}p_{y} \\
        & = v_{F}p_{x}\kappa^{2}i\hbar\sigma_{y} - v_{F}p_{y}\kappa i\hbar\sigma_{x}.
    \end{split}
\end{equation}

Calculating the Hamiltonian commutator again with the sum of the angular momentum and the lattice spin operator, we have,

\begin{equation}
    \begin{split}
        [\mathcal{H'}, (\mathbf{L}_{z} + \mathbf{S}_{z})] & = [\mathcal{H'}, \mathbf{L}_{z}] + [\mathcal{H'}, \mathbf{S}_{z}] \\
        & = (-v_{F}\kappa\sigma_{x}i\hbar p_{y} + v_{F}\sigma_{y}i\hbar p_{x}) \\
        & + (v_{F}p_{x}\kappa^{2}i\hbar\sigma_{y} - v_{F}p_{y}\kappa i\hbar\sigma_{x}) \\
        & = i\hbar v_{F}\kappa^2\sigma_{y}p_{x}-i\hbar v_{F}\sigma_{y}p_{x} \\
        & = i\hbar v_{F}\sigma_{y}p_{x}(\kappa^2-1);
    \end{split}
\end{equation}

\noindent remembering that \(\kappa = \pm 1\) can have only these two values, the result of our commutator will be, 

\begin{equation}
    [\mathcal{H'}, (\mathbf{L}_{z} + \mathbf{S}_{z})] = 0,
\end{equation}

\noindent checked that \([\mathcal{H'}, (\mathbf{L}_{z} + \mathbf{S}_{z})] = 0\) for any value of \(\Delta\) and \(\delta\) holds.

In this way, we were able to generate a time-reversal symmetry breaking between the Dirac cones of this honeycomb lattice.

The complete Hamiltonian, which describes both pairs of Dirac cones ($K$ and $K'$) and includes our engineered symmetry-breaking term, is given in \autoref{eq:hamiltonian-pairs-combined}. In this matrix, each non-zero block represents one of the valleys. The energy gap, modified by the $\delta$ term, appears on the main diagonal. This diagonal placement represents the different on-site energies of the sublattices, a feature essential for modeling the hopping processes between neighboring sites.

This diagonal placement allows for a clear representation of how the energy varies across different sites in the lattice, which is essential for modeling hopping processes between neighboring sites.

With the theoretical framework now established in this chapter, we have successfully formulated a physically robust Hamiltonian. This model overcomes the limitations of pristine graphene by explicitly breaking sublattice symmetry to introduce a mass gap ($\Delta$) and, crucially, incorporates a time-reversal symmetry-breaking parameter ($\delta$) to engineer a controllable valley imbalance. 

The validation of this model, through the confirmation that the total angular momentum (orbital plus lattice spin) remains a conserved quantity, provides a consistent foundation for the next and most critical step: investigating the response of the system to an external magnetic field to probe for a CME analogue.

\section{Coupling the Magnetic Field}

As established in Chapter 4, the theoretical model is now complete and validated. We have successfully formulated a Hamiltonian that not only breaks the sublattice symmetry of the honeycomb lattice to create a gapped system but also introduces a valley imbalance through a time-reversal symmetry-breaking term, $\delta$. This engineered imbalance is the direct analogue to the chiral chemical potential required for the CME. The next logical and most critical step in this investigation is to introduce an external magnetic field to this system and analyze its response.

The complete low-energy Hamiltonian, accounting the spin of electron\footnote{Spin up and spin down} (represented by the two $4\times4$ blocks), the two sublattices (within each $2\times2$ block), and the two valleys ($K$ and $K'$) with their imbalanced gaps ($m\pm\delta$), is given by:

\begin{equation}
    \scalebox{0.6}{$\mathcal{H} = \left(\begin{array}{cccc|cccc}
    m+\delta & v_{F}(p_{x}-ip_{y}) & 0 & 0 & 0 & 0 & 0 & 0 \\
    v_{F}(p_{x}+ip_{y}) & -m-\delta & 0 & 0 & 0 & 0 & 0 & 0 \\
    0 & 0 & m-\delta & v_{F}(-p_{x}-ip_{y}) & 0 & 0 & 0 & 0 \\
    0 & 0 & v_{F}(-p_{x}+ip_{y}) & -m+\delta & 0 & 0 & 0 & 0 \\
    \hline
    0 & 0 & 0 & 0 & -m-\delta & v_{F}(p_{x}-ip_{y}) & 0 & 0 \\
    0 & 0 & 0 & 0 & v_{F}(p_{x}+ip_{y}) & m+\delta & 0 & 0 \\
    0 & 0 & 0 & 0 & 0 & 0 & -m+\delta & v_{F}(-p_{x}-ip_{y}) \\
    0 & 0 & 0 & 0 & 0 & 0 & v_{F}(-p_{x}+ip_{y}) & m-\delta
    \end{array}\right)$}.
\end{equation}

Coupling an in-plane magnetic field to a quasi-$2D$ system like this one presents a subtle but important choice in theoretical modeling: the selection of the vector potential gauge. While different gauges must ultimately yield the same physical predictions, they can lead to Hamiltonians that appear mathematically distinct and offer different physical insights. A detailed analysis reveals the equivalence and specific physical interpretation of the two primary gauges used for this purpose.

The analysis of how a uniform in-plane magnetic field affects the electronic Hamiltonian of graphene, which is modeled using a 4-component spinor basis that accounts for valley ($K,K'$) and sublattice ($A,B$) degrees of freedom, is key here. Two different vector potential gauges are considered, $A^{(y)}=(0,-zB_{x},0)$ and $A^{(z)}=(0,0,yB_{x})$, both of which generate the same magnetic field. The primary conclusion for an idealized, perfectly flat graphene sheet at $z=0$ is that there is no orbital effect from the in-plane field. In this strictly $2D$ limit, minimal coupling ($\mathbf{k} \mapsto \mathbf{k} - \frac{q}{\hbar}\mathbf{A}$) results in no change to the in-plane Hamiltonian in either gauge, leaving only the Zeeman term to couple the field to the electron real spin.

The analysis then transitions to the more physically realistic case where the graphene sheet has a finite vertical extent, such as ripples or sublattices at different z-positions; for this, we focus exclusively on the orbital effects and ignore the Zeeman interaction. In this scenario, orbital effects do emerge, and it can be demonstrated how two different gauges provide equivalent descriptions of the same physics. In the first gauge, using the vector potential $\vec{A}^{(y)}=(0,-zB_x,0)$, the interaction is introduced via minimal coupling, which, when averaged over the vertical wavefunction of the electron, results in a simple momentum shift in the effective $2D$ Hamiltonian where $k_{y}$ is replaced by $k_{y} + \frac{q}{\hbar}\bar{z}B_{x}$, with $\bar{z}$ being the average vertical position. In the second gauge, using the vector potential $\vec{A}^{(z)}=(0,0,yB_x)$, the in-plane components of the vector potential are zero, and the effect manifests differently: the hopping terms of the Hamiltonian are modified by a position-dependent phase factor, $\exp\left(i\frac{q}{\hbar}yB_x\Delta z\right)$, which is acquired by the wavefunction of electron during any movement over a vertical distance $\Delta z$. The two descriptions are gauge-equivalent, as the constant momentum shift in the first gauge is the direct result of the Fourier transform of the position-dependent phase acquired during hopping in the second.

Following this understanding, we will explore two distinct approaches to couple a magnetic field to this system. Both approaches use gauges that produce an in-plane magnetic field ($B\hat{x}$), a configuration motivated by the idea that the quasi-planar nature of the material, with its finite extent in the z-direction, might be sufficient to enable a CME-like effect.

\subsubsection{Approach 1: In-Plane Field via Minimal Coupling}

The first approach applies the magnetic field using a vector potential [\autoref{appendice:d}] oriented in the y-direction:

\begin{equation}
    \vec{A} = (0, -zB, 0), \quad (\text{which gives } \vec{B} = B\hat{x}).
\end{equation}

This is done via minimal coupling, where the momentum operator $\vec{p}$ is replaced by $\vec{p} - q\vec{A}$. Applying this substitution modifies the off-diagonal kinetic terms of the Hamiltonian. For the spin-up components, the resulting Hamiltonians for the two valleys are:

\noindent $K'_{\uparrow}$ Valley:
\begin{equation}
    \scalebox{0.8}{$\mathcal{H}_{K'_{\uparrow}} =
\begin{pmatrix}
 m+\delta & v_{F}\left(\left(k_{x}-\frac{2\pi}{3a}\right)-i\left(\left(k_{y}-\frac{2\pi}{3\sqrt{3}a}\right) + \left(\frac{qBz}{c}\right)\right)\right) \\
 v_{F}\left(\left(k_{x}-\frac{2\pi}{3a}\right)+i\left(\left(k_{y}-\frac{2\pi}{3\sqrt{3}a}\right) + \left(\frac{qBz}{c}\right)\right)\right) & -m-\delta
\end{pmatrix}$},
\end{equation}

\noindent $K_{\uparrow}$ Valley:

\begin{equation}
    \scalebox{0.8}{$\mathcal{H}_{K_{\uparrow}} =
    \begin{pmatrix}
    m-\delta & v_{F}\left(-\left(k_{x}-\frac{2\pi}{3a}\right)-i\left(\left(k_{y}+\frac{2\pi}{3\sqrt{3}a}\right)+\left(\frac{qBz}{c}\right)\right)\right)\\
    v_{F}\left(-\left(k_{x}-\frac{2\pi}{3a}\right)+i\left(\left(k_{y}+\frac{2\pi}{3\sqrt{3}a}\right)+\left(\frac{qBz}{c}\right)\right)\right) & -m+\delta
    \end{pmatrix}$}.
\end{equation}

\subsubsection{Approach 2: In-Plane Field via Diagonal Coupling}\label{second-approach}

The second approach uses a vector potential oriented in the z-direction:

\begin{equation}
    \vec{A} = (0, 0, yB), \quad (\text{which also gives } \vec{B} = B\hat{x})
\end{equation}

In this case, the interaction is introduced differently, modifying the diagonal mass terms of the Hamiltonian directly. The resulting $8\times8$ Hamiltonian is:

\begin{equation}
    \scalebox{0.54}{$\mathcal{H} = \left(\begin{array}{cccc|cccc}
m+\delta-\left(\frac{qBy}{c}\right) & v_{F}(p_{x}-ip_{y}) & 0 & 0 & 0 & 0 & 0 & 0 \\
v_{F}(p_{x}+ip_{y}) & -m-\delta+\left(\frac{qBy}{c}\right) & 0 & 0 & 0 & 0 & 0 & 0 \\
0 & 0 & m-\delta+\left(\frac{qBy}{c}\right) & v_{F}(-p_{x}-ip_{y}) & 0 & 0 & 0 & 0 \\
0 & 0 & v_{F}(-p_{x}+ip_{y}) & -m+\delta-\left(\frac{qBy}{c}\right) & 0 & 0 & 0 & 0 \\
\hline
0 & 0 & 0 & 0 & -m-\delta+\left(\frac{qBy}{c}\right) & v_{F}(p_{x}-ip_{y}) & 0 & 0 \\
0 & 0 & 0 & 0 & v_{F}(p_{x}+ip_{y}) & m+\delta-\left(\frac{qBy}{c}\right) & 0 & 0 \\
0 & 0 & 0 & 0 & 0 & 0 & -m+\delta-\left(\frac{qBy}{c}\right) & v_{F}(-p_{x}-ip_{y}) \\
0 & 0 & 0 & 0 & 0 & 0 & v_{F}(-p_{x}+ip_{y}) & m-\delta+\left(\frac{qBy}{c}\right)
\end{array}\right)$}
\end{equation}

\subsection{Final proposal}

Adopting the second approach [subsection \ref{second-approach}], where the in-plane magnetic field is introduced via the vector potential $\vec{A}=(0,0,yB)$, we can reformulate the resulting Hamiltonian in a more general, coordinate-independent notation. It is important to emphasize that this choice is made for mathematical convenience, as both approaches are physically equivalent due to gauge invariance. Both vector potentials, $A^{(y)}=(0,-zB,0)$ and $A^{(z)}=(0,0,yB)$, generate the identical magnetic field and are related by a gauge transformation. Their effect on the Hamiltonian simply manifests in different ways: the first gauge produces a constant shift in the kinetic momentum ($k_y$), while the second introduces a position-dependent phase to the hopping terms between vertically-separated sites. These descriptions are equivalent, as the momentum shift is the Fourier transform of the position-dependent phase.

We proceed with the second approach because it is particularly elegant for our theoretical framework. This method recasts the magnetic interaction as a position-dependent modification to the mass term of the system. This allows the effect to be seamlessly absorbed into our coordinate-independent formalism. To capture this structure, we define a generalized vector $\mathbf{u''}$ that incorporates the kinetic, intrinsic mass, and magnetic field components of the system.

This generalized vector will be defined as:

\begin{equation}
    \mathbf{u''} \equiv (\mathbf{p}\cdot\hat{\mathbf{a}}_{d})\hat{\mathbf{a}}_{d} + (\mathbf{p}\cdot\hat{\mathbf{a}}_s)\hat{\mathbf{a}}_{s} + \kappa(\Delta/v_{F}+\kappa\delta-\kappa\frac{qBy}{c})\hat{\mathbf{a}}_{n}.
\end{equation}

Here, the contribution of the magnetic field, $-\kappa\frac{qBy}{c}$, is absorbed directly into the normal component ($\hat{\mathbf{a}}_{n}$), which physically corresponds to the effective mass. This powerful formulation allows the complete Hamiltonian, now including the magnetic coupling, to be expressed in the remarkably compact and elegant form:

\begin{equation}
    \mathcal{H''} = \frac{2v_{F}}{\hbar}\mathbf{S}\cdot\mathbf{u''}.
\end{equation}
 
This compact form of the Hamiltonian provides a robust foundation for a detailed analysis of the transport properties of the systems. It establishes a framework for future calculations based on linear response theory \cite{Dudal:2021ret} or other approaches, enabling the exploration of the conductivity and topological characteristics of the systems. The theoretical setup was specifically engineered to mimic the symmetry pattern of the anomalous quark-gluon plasma by introducing a controllable imbalance in pseudo-chirality. The ultimate goal is to determine if coupling the magnetic field to the emergent lattice spin can generate an electrical current parallel to the applied field.

A central consideration for this effect is the dimensionality of the system. While an in-plane magnetic field has a null orbital effect on a strictly two-dimensional system, real-world materials are never perfectly flat. Considering a small but finite thickness in the z-direction allows for a non-trivial coupling that may give rise to electric currents driven by the magnetic field. This can be understood in terms of the representations of the SU(2) group that will change depending on the dimensionality of the system: for strictly $2D$ systems, the group representation involves only one rotation (around an axis perpendicular to the plane) and two boosts. However, in a $3D$ context, the three Pauli matrices correspond to three independent rotations. We argue that the unavoidable three-dimensional character of any physical material is precisely what enables this phenomenon.

The perspective is that the work done in this thesis will benefit future investigations. The next logical step is to calculate the transport properties of this system using the established Hamiltonian. This will allow for the identification of good candidates for the physical realization of our setup and a definitive answer as to whether a net anomalous current can be generated.

\chapter{Conclusion}
This dissertation undertook a systematic theoretical investigation into the possibility of realizing an analogue of the CME in a quasi-planar, two-dimensional condensed matter system. The foundational premise of this work was to establish a rigorous conceptual bridge between the physics of anomalous transport in high-energy relativistic systems, such as the QGP, and the emergent electronic phenomena within engineered honeycomb lattices. The central research objective was to formulate a Hamiltonian that not only captures the essential features of a $2+1D$ system with relativistic-like quasiparticles but also incorporates the necessary ingredients for a CME-like phenomenon: a well-defined pseudo-chirality and a mechanism for inducing a non-equilibrium imbalance between the corresponding pseudo-chiral populations.

The investigation commenced with a thorough review of the theoretical underpinnings of the CME in its native $3+1D$ context, identifying massless chiral fermions and a finite chiral chemical potential as the indispensable components. This foundational understanding established the precise criteria that any condensed matter analogue must satisfy. The subsequent analysis of pristine graphene confirmed its status as a promising but ultimately incomplete platform. While its low-energy excitations manifest as massless Dirac fermions within two distinct valleys — providing a natural analogue to chirality — its inherent sublattice symmetry, protected by the identical nature of its carbon atoms, precludes the opening of a mass gap. This gapless nature, a celebrated feature of graphene, proves to be a fundamental impediment in this context, as it prevents the definition of robust, distinct pseudo-chiral states necessary to replicate the conditions for the CME.

The primary theoretical contribution of this work was the development of a model that overcomes this fundamental limitation. By positing a honeycomb lattice with explicitly broken sublattice symmetry — a condition achievable by constituting the lattice from two distinct atomic species — a mass term was introduced into the low-energy effective Hamiltonian. This term, proportional to the Pauli matrix $\sigma_{z}$, opens a band gap and fundamentally alters the quantum mechanics of the system. It endows the system with a new, well-defined quantum number, termed pseudo-chirality, which is directly related to the sublattice polarization of the charge carriers. This emergent property, where eigenstates have a higher probability of residing on one sublattice over the other, serves as a precise and powerful analogue to the chirality of relativistic fermions.

Building upon this gapped system, the concept of ``lattice spin'' — an emergent form of angular momentum arising from the interaction between charge carriers and the underlying lattice structure — was incorporated into the model. The critical step toward engineering a CME-like state was the introduction of a time-reversal symmetry-breaking parameter, $\delta$, into the Hamiltonian. This parameter was designed to asymmetrically modify the energy gap in the two non - equivalent valleys, thereby creating a controllable imbalance in the valley populations. This engineered imbalance, which creates an excess of carriers in one valley over the other, functions as a direct analogue to the chiral chemical potential ($\mu_5$) that drives the CME in the QGP and Weyl semimetals. It represents the successful translation of a key high-energy concept into a controllable parameter within a condensed matter Hamiltonian.

A significant outcome of this research was the validation of the internal consistency and physical robustness of the model. Through explicit commutator calculations, it was demonstrated that the sum of the orbital angular momentum and the lattice spin remains a conserved quantity along the axis of rotational symmetry, even in the presence of the symmetry-breaking terms ($\Delta$ and $\delta$). This non-trivial result is of critical importance, as it confirms that the proposed Hamiltonian, despite its modifications to introduce new physics, respects fundamental conservation laws. This provides a robust and physically consistent foundation for describing the dynamics of charge carriers in such a system and lends significant credence to the model as a viable representation of a physical system.

With the theoretical model established and validated, the final and most critical step was to introduce an external magnetic field to this system and analyze its response. This was achieved by exploring two distinct approaches to couple an in-plane magnetic field, a configuration motivated by the material quasi-planar nature and finite extent in the z-direction. The first approach applied the field via minimal coupling using a vector potential $\vec{A}=(0,-zB,0)$, whereas the second approach used a vector potential $\vec{A}=(0,0,yB)$ to modify the diagonal mass terms of the Hamiltonian directly. Adopting this second approach, the research culminated in a final proposal that reformulates the Hamiltonian in a remarkably compact, coordinate-independent form, $\mathcal{H''} = \frac{2v_{F}}{\hbar}\mathbf{S}\cdot\mathbf{u''}$, which elegantly incorporates the kinetic, mass, valley imbalance, and magnetic field components. This final Hamiltonian now serves as the foundation for the subsequent analysis required to calculate the transport properties and determine if a net anomalous current is generated.

In conclusion, this dissertation has successfully constructed a self-consistent theoretical framework that establishes the necessary conditions for observing a CME analogue in a $2+1D$ honeycomb lattice. The work demonstrates that the explicit breaking of sublattice symmetry is the essential first step, creating a gapped system wherein pseudo-chirality becomes a well-defined quantum number based on sublattice polarization. It further shows that a non-equilibrium imbalance between these pseudo-chiral states — a valley polarization — can be controllably engineered through the introduction of a carefully constructed time-reversal symmetry-breaking term. The validation of the adherence of the model to the conservation of total angular momentum solidifies its theoretical integrity. By systematically assembling these components and coupling the system to an external magnetic field, this research has transformed the abstract possibility of a $2D$ CME into a concrete theoretical model, thereby completing the foundational task of building a viable conceptual bridge between two distinct and fundamental areas of physics.
\selectlanguage{english}
\typeout{}
\bibliographystyle{unsrt}
\bibliography{references}

@article{PhysRevX.5.031013,
  title = {Experimental Discovery of Weyl Semimetal TaAs},
  author = {Lv, B. Q. and Weng, H. M. and Fu, B. B. and Wang, X. P. and Miao, H. and Ma, J. and Richard, P. and Huang, X. C. and Zhao, L. X. and Chen, G. F. and Fang, Z. and Dai, X. and Qian, T. and Ding, H.},
  journal = {Phys. Rev. X},
  volume = {5},
  issue = {3},
  pages = {031013},
  numpages = {8},
  year = {2015},
  month = {Jul},
  publisher = {American Physical Society},
  doi = {10.1103/PhysRevX.5.031013},
  url = {https://link.aps.org/doi/10.1103/PhysRevX.5.031013}
}

@article{Li2016,
  author    = {Li, Qiang and Kharzeev, Dmitri E. and Zhang, Cheng and Huang, Yuan and Pletikosić, I. and Fedorov, A. V. and Zhong, R. D. and Schneeloch, J. A. and Gu, G. D. and Valla, T.},
  journal   = {Nature Physics},
  title     = {Chiral magnetic effect in ZrTe5},
  year      = {2016},
  issn      = {1745-2481},
  month     = feb,
  number    = {6},
  pages     = {550--554},
  volume    = {12},
  doi       = {10.1038/nphys3648},
  publisher = {Springer Science and Business Media LLC},
}

@article{KHARZEEV2008227,
    title = {The effects of topological charge change in heavy ion collisions: “Event by event P and CP violation”},
    journal = {Nuclear Physics A},
    volume = {803},
    number = {3},
    pages = {227-253},
    year = {2008},
    issn = {0375-9474},
    doi = {https://doi.org/10.1016/j.nuclphysa.2008.02.298},
    url = {https://www.sciencedirect.com/science/article/pii/S037594740800078X},
    author = {Dmitri E. Kharzeev and Larry D. McLerran and Harmen J. Warringa}
}

@article{PhysRevC.105.014901,
  title = {Search for the chiral magnetic effect with isobar collisions at $\sqrt{{s}_{NN}}=200$ GeV by the STAR Collaboration at the BNL Relativistic Heavy Ion Collider},
  author = {M. S. Abdallah et al. (STAR Collaboration)},
  collaboration = {STAR Collaboration},
  journal = {Phys. Rev. C},
  volume = {105},
  issue = {1},
  pages = {014901},
  numpages = {34},
  year = {2022},
  month = {Jan},
  publisher = {American Physical Society},
  doi = {10.1103/PhysRevC.105.014901},
  url = {https://link.aps.org/doi/10.1103/PhysRevC.105.014901}
}

@article{PhysRevX.5.031023,
  title = {Observation of the Chiral-Anomaly-Induced Negative Magnetoresistance in 3D Weyl Semimetal TaAs},
  author = {Huang, Xiaochun and Zhao, Lingxiao and Long, Yujia and Wang, Peipei and Chen, Dong and Yang, Zhanhai and Liang, Hui and Xue, Mianqi and Weng, Hongming and Fang, Zhong and Dai, Xi and Chen, Genfu},
  journal = {Phys. Rev. X},
  volume = {5},
  issue = {3},
  pages = {031023},
  numpages = {9},
  year = {2015},
  month = {Aug},
  publisher = {American Physical Society},
  doi = {10.1103/PhysRevX.5.031023},
  url = {https://link.aps.org/doi/10.1103/PhysRevX.5.031023}
}

@article{PhysRevB.93.115414,
  title = {Transport evidence for the three-dimensional Dirac semimetal phase in $\mathrm{ZrT}{\mathrm{e}}_{5}$},
  author = {Zheng, Guolin and Lu, Jianwei and Zhu, Xiangde and Ning, Wei and Han, Yuyan and Zhang, Hongwei and Zhang, Jinglei and Xi, Chuanying and Yang, Jiyong and Du, Haifeng and Yang, Kun and Zhang, Yuheng and Tian, Mingliang},
  journal = {Phys. Rev. B},
  volume = {93},
  issue = {11},
  pages = {115414},
  numpages = {7},
  year = {2016},
  month = {Mar},
  publisher = {American Physical Society},
  doi = {10.1103/PhysRevB.93.115414},
  url = {https://link.aps.org/doi/10.1103/PhysRevB.93.115414}
}

@article{PhysRevLett.106.116803,
  title = {Spin and the Honeycomb Lattice: Lessons from Graphene},
  author = {Mecklenburg, Matthew and Regan, B. C.},
  journal = {Phys. Rev. Lett.},
  volume = {106},
  issue = {11},
  pages = {116803},
  numpages = {4},
  year = {2011},
  month = {Mar},
  publisher = {American Physical Society},
  doi = {10.1103/PhysRevLett.106.116803},
  url = {https://link.aps.org/doi/10.1103/PhysRevLett.106.116803}
}

@book{graphene,
    author = {Mikhail I. Katsnelson},
    title = {Graphene: Carbon in Two Dimensions},
    publisher = {Cambridge University Press},
    year = {2012}
}

@article{PhysRevLett.103.251601,
  title = {Azimuthal Charged-Particle Correlations and Possible Local Strong Parity Violation},
  author = {Abelev, B. I. and Aggarwal, M. M. and Ahammed, Z. and Alakhverdyants, A. V. and Anderson, B. D. and Arkhipkin, D. and Averichev, G. S. and Balewski, J. and Barannikova et al. (STAR Collaboration), O.},
  collaboration = {STAR Collaboration},
  journal = {Phys. Rev. Lett.},
  volume = {103},
  issue = {25},
  pages = {251601},
  numpages = {7},
  year = {2009},
  month = {Dec},
  publisher = {American Physical Society},
  doi = {10.1103/PhysRevLett.103.251601},
  url = {https://link.aps.org/doi/10.1103/PhysRevLett.103.251601}
}

@misc{pratt2010alternativecontributionsangularcorrelations,
      title={Alternative Contributions to the Angular Correlations Observed at RHIC Associated with Parity Fluctuations}, 
      author={Scott Pratt},
      year={2010},
      eprint={1002.1758},
      archivePrefix={arXiv},
      primaryClass={nucl-th},
      url={https://arxiv.org/abs/1002.1758}
}

@article{PhysRevB.83.205101,
  title = {Topological semimetal and Fermi-arc surface states in the electronic structure of pyrochlore iridates},
  author = {Wan, Xiangang and Turner, Ari M. and Vishwanath, Ashvin and Savrasov, Sergey Y.},
  journal = {Phys. Rev. B},
  volume = {83},
  issue = {20},
  pages = {205101},
  numpages = {9},
  year = {2011},
  month = {May},
  publisher = {American Physical Society},
  doi = {10.1103/PhysRevB.83.205101},
  url = {https://link.aps.org/doi/10.1103/PhysRevB.83.205101}
}

@article{PhysRevLett.107.127205,
  title = {Weyl Semimetal in a Topological Insulator Multilayer},
  author = {Burkov, A. A. and Balents, Leon},
  journal = {Phys. Rev. Lett.},
  volume = {107},
  issue = {12},
  pages = {127205},
  numpages = {4},
  year = {2011},
  month = {Sep},
  publisher = {American Physical Society},
  doi = {10.1103/PhysRevLett.107.127205},
  url = {https://link.aps.org/doi/10.1103/PhysRevLett.107.127205}
}

@article{PhysRevB.86.115133,
  title = {Topological response in Weyl semimetals and the chiral anomaly},
  author = {Zyuzin, A. A. and Burkov, A. A.},
  journal = {Phys. Rev. B},
  volume = {86},
  issue = {11},
  pages = {115133},
  numpages = {8},
  year = {2012},
  month = {Sep},
  publisher = {American Physical Society},
  doi = {10.1103/PhysRevB.86.115133},
  url = {https://link.aps.org/doi/10.1103/PhysRevB.86.115133}
}

@Article{Li2013,
  author    = {Li, Xiao and Cao, Ting and Niu, Qian and Shi, Junren and Feng, Ji},
  journal   = {Proceedings of the National Academy of Sciences},
  title     = {Coupling the valley degree of freedom to antiferromagnetic order},
  year      = {2013},
  issn      = {1091-6490},
  month     = feb,
  number    = {10},
  pages     = {3738--3742},
  volume    = {110},
  doi       = {10.1073/pnas.1219420110},
  publisher = {Proceedings of the National Academy of Sciences},
}

@article{Fukushima:2008xe,
    author = "Fukushima, Kenji and Kharzeev, Dmitri E. and Warringa, Harmen J.",
    title = "{The Chiral Magnetic Effect}",
    eprint = "0808.3382",
    archivePrefix = "arXiv",
    primaryClass = "hep-ph",
    doi = "10.1103/PhysRevD.78.074033",
    journal = "Phys. Rev. D",
    volume = "78",
    pages = "074033",
    year = "2008"
}

@article{Katsnelson_2006,
   title={Chiral tunnelling and the Klein paradox in graphene},
   volume={2},
   ISSN={1745-2481},
   url={http://dx.doi.org/10.1038/nphys384},
   DOI={10.1038/nphys384},
   number={9},
   journal={Nature Physics},
   publisher={Springer Science and Business Media LLC},
   author={Katsnelson, M. I. and Novoselov, K. S. and Geim, A. K.},
   year={2006},
   month=aug, pages={620–625} }

@article{VOZMEDIANO2010109,
    title = {Gauge fields in graphene},
    journal = {Physics Reports},
    volume = {496},
    number = {4},
    pages = {109-148},
    year = {2010},
    issn = {0370-1573},
    doi = {https://doi.org/10.1016/j.physrep.2010.07.003},
    url = {https://www.sciencedirect.com/science/article/pii/S0370157310001729},
    author = {M.A.H. Vozmediano and M.I. Katsnelson and F. Guinea}
}

@article{Cortijo_2012,
   title={Geometrical and topological aspects of graphene and related materials},
   volume={45},
   ISSN={1751-8121},
   url={http://dx.doi.org/10.1088/1751-8113/45/38/383001},
   DOI={10.1088/1751-8113/45/38/383001},
   number={38},
   journal={Journal of Physics A: Mathematical and Theoretical},
   publisher={IOP Publishing},
   author={Cortijo, A and Guinea, F and Vozmediano, M A H},
   year={2012},
   month=sep, pages={383001} }

@article{Chernodub_2014,
   title={Condensed matter realization of the axial magnetic effect},
   volume={89},
   ISSN={1550-235X},
   url={http://dx.doi.org/10.1103/PhysRevB.89.081407},
   DOI={10.1103/physrevb.89.081407},
   number={8},
   journal={Physical Review B},
   publisher={American Physical Society (APS)},
   author={Chernodub, Maxim N. and Cortijo, Alberto and Grushin, Adolfo G. and Landsteiner, Karl and Vozmediano, María A. H.},
   year={2014},
   month=feb }

@article{Iorio_2012,
   title={The Hawking–Unruh phenomenon on graphene},
   volume={716},
   ISSN={0370-2693},
   url={http://dx.doi.org/10.1016/j.physletb.2012.08.023},
   DOI={10.1016/j.physletb.2012.08.023},
   number={2},
   journal={Physics Letters B},
   publisher={Elsevier BV},
   author={Iorio, Alfredo and Lambiase, Gaetano},
   year={2012},
   month=sep, pages={334–337} }

@article{Iorio_2014,
   title={Quantum field theory in curved graphene spacetimes, Lobachevsky geometry, Weyl symmetry, Hawking effect, and all that},
   volume={90},
   ISSN={1550-2368},
   url={http://dx.doi.org/10.1103/PhysRevD.90.025006},
   DOI={10.1103/physrevd.90.025006},
   number={2},
   journal={Physical Review D},
   publisher={American Physical Society (APS)},
   author={Iorio, Alfredo and Lambiase, Gaetano},
   year={2014},
   month=jul 
}

@book{Cheng:1984vwu,
    author = "Cheng, Ta-Pei [0000-0002-1137-0969] and Li, Ling-Fong [0000-0002-8035-3329]",
    title = "{Gauge Theory of Elementary Particle Physics}",
    isbn = "978-0-19-851961-4, 978-0-19-851961-4",
    publisher = "Oxford University Press",
    address = "Oxford, UK",
    year = "1984"
}

@misc{behnami2025chiralanomalyweylsemimetal,
      title={Chiral anomaly in the Weyl semimetal TaRhTe$_4$}, 
      author={M. Behnami and D. V. Efremov and S. Aswartham and G. Shipunov and B. R. Piening and C. G. F. Blum and V. Kocsis and J. Dufouleur and I. Pallecchi and M. Putti and B. Büchner and H. Reichlova and F. Caglieris},
      year={2025},
      eprint={2502.18937},
      archivePrefix={arXiv},
      primaryClass={cond-mat.str-el},
      url={https://arxiv.org/abs/2502.18937}, 
}

@article{Nielsen:1981hk,
    author = "Nielsen, Holger Bech and Ninomiya, M.",
    title = "{No Go Theorem for Regularizing Chiral Fermions}",
    reportNumber = "RL-81-052",
    doi = "10.1016/0370-2693(81)91026-1",
    journal = "Phys. Lett. B",
    volume = "105",
    pages = "219--223",
    year = "1981"
}

@article{NIELSEN1983389,
    title = {The Adler-Bell-Jackiw anomaly and Weyl fermions in a crystal},
    journal = {Physics Letters B},
    volume = {130},
    number = {6},
    pages = {389-396},
    year = {1983},
    issn = {0370-2693},
    doi = {https://doi.org/10.1016/0370-2693(83)91529-0},
    url = {https://www.sciencedirect.com/science/article/pii/0370269383915290},
    author = {H.B. Nielsen and Masao Ninomiya}
}

@article{Dudal:2021ret,
    author = "Dudal, David and Matusalem, Filipe and Mizher, Ana J{\'u}lia and Rocha, Alexandre Reily and Villavicencio, Cristian",
    title = "{Half-integer anomalous currents in 2D materials from a QFT viewpoint}",
    eprint = "2103.10341",
    archivePrefix = "arXiv",
    primaryClass = "cond-mat.mes-hall",
    doi = "10.1038/s41598-022-09483-4",
    journal = "Sci. Rep.",
    volume = "12",
    number = "1",
    pages = "5439",
    year = "2022"
}

@article{Mizher:2018dtf,
    author = "Mizher, Ana Julia and Hernandez-Ortiz, Saul and Raya, Alfredo and Villavicencio, Cristian",
    title = "{Aspects of the pseudo Chiral Magnetic Effect in 2D Weyl-Dirac Matter}",
    eprint = "1803.05794",
    archivePrefix = "arXiv",
    primaryClass = "hep-ph",
    doi = "10.1140/epjc/s10052-018-6380-1",
    journal = "Eur. Phys. J. C",
    volume = "78",
    number = "11",
    pages = "912",
    year = "2018"
}

@article{PhysRevD.44.3501,
  title = {hijing: A Monte Carlo model for multiple jet production in $\mathrm{pp}$, $\mathrm{pA}$, and $\mathrm{AA}$ collisions},
  author = {Wang, Xin-Nian and Gyulassy, Miklos},
  journal = {Phys. Rev. D},
  volume = {44},
  issue = {11},
  pages = {3501--3516},
  numpages = {0},
  year = {1991},
  month = {Dec},
  publisher = {American Physical Society},
  doi = {10.1103/PhysRevD.44.3501},
  url = {https://link.aps.org/doi/10.1103/PhysRevD.44.3501}
}

@article{MBleicher_1999,
    doi = {10.1088/0954-3899/25/9/308},
    url = {https://doi.org/10.1088/0954-3899/25/9/308},
    year = {1999},
    month = {sep},
    publisher = {},
    volume = {25},
    number = {9},
    pages = {1859},
    author = {M Bleicher and E Zabrodin and C Spieles and S A Bass and C Ernst and S Soff and L Bravina and M Belkacem and H Weber and H Stöcker and W Greiner},
    title = {Relativistic 
    hadron-hadron collisions in the ultra-relativistic 
    quantum molecular dynamics model},
    journal = {Journal of Physics G: Nuclear and Particle Physics}
}

@misc{ray2000mevsimmontecarloevent,
      title={MEVSIM: A Monte Carlo Event Generator for STAR}, 
      author={R. L. Ray and R. S. Longacre},
      year={2000},
      eprint={nucl-ex/0008009},
      archivePrefix={arXiv},
      primaryClass={nucl-ex},
      url={https://arxiv.org/abs/nucl-ex/0008009}, 
}

@article{Adhikari:2024bfa,
    author = "Adhikari, Prabal and others",
    title = "{Strongly interacting matter in extreme magnetic fields}",
    eprint = "2412.18632",
    archivePrefix = "arXiv",
    primaryClass = "nucl-th",
    doi = "10.1016/j.ppnp.2025.104199",
    journal = "Prog. Part. Nucl. Phys.",
    volume = "146",
    pages = "104199",
    year = "2026"
}

%
%


\begin{apendicesenv}


\chapter{Reciprocal Lattice}
\label{appendix:a}
The reciprocal lattice [\ref{fig:reciprocal-space}] is a concept in solid-state physics and crystallography that provides a mathematical framework for understanding the behavior of waves (such as electromagnetic waves or quantum mechanical wave functions) in periodic structures such as crystals. It is a crucial tool for understanding various phenomena in condensed matter physics, including the behavior of electrons in a crystal lattice, the diffraction of x-rays by crystals, and the formation of electronic band structures.

\begin{figure}[htbp!]
    \centering
    \includegraphics[width=0.8\linewidth]{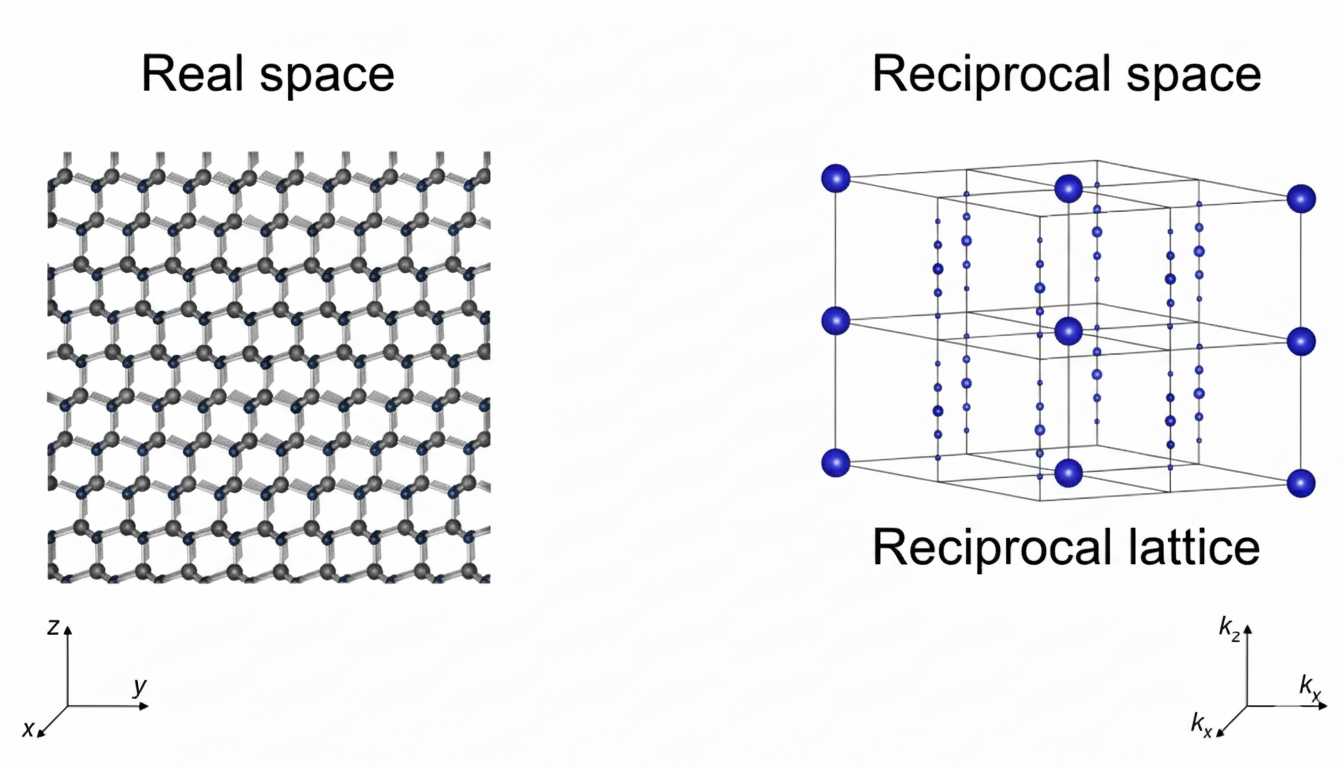}
    \caption{Real and reciprocal lattice}
    \label{fig:reciprocal-space}
\end{figure}

The coordinates of the real space we live in are \(x, y, z\). On the other hand, the space where the coordinates are \(k_{x}, k_{y}, k_{z}\) is the reciprocal space. The lattice in the reciprocal space is called the reciprocal lattice. There are two reasons why we have to think about the reciprocal lattice. The first is to know the crystal structure of the material.

In order to know the atomic arrangement of matter, we need to perform an experiment called diffraction [\ref{fig:diffraction-experiment}]. The diffraction experiment is a method for observing the reciprocal lattice. By analyzing the observed reciprocal lattice, the crystal structure corresponding to the real lattice can be known.

\begin{figure}[htbp!]
    \centering
    \includegraphics[width=0.8\linewidth]{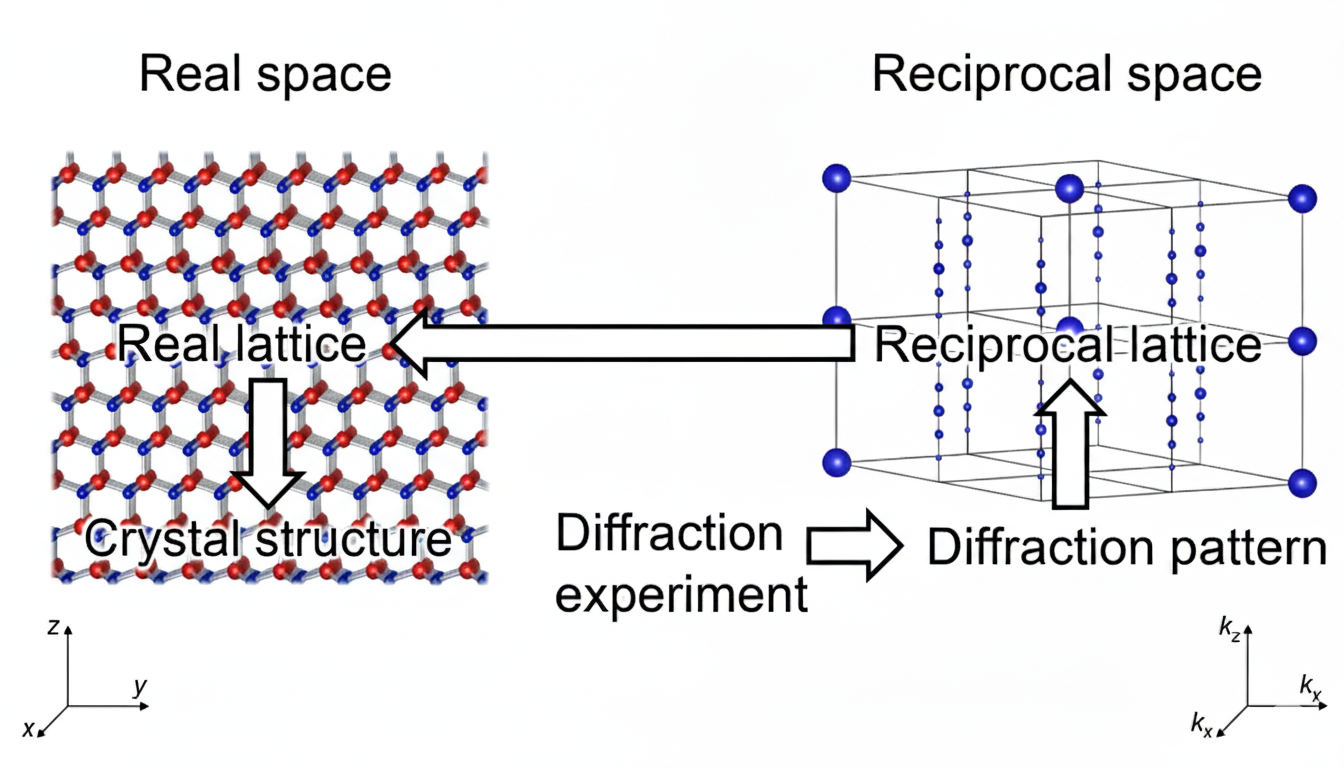}
    \caption{Diffraction experiment }
    \label{fig:diffraction-experiment}
\end{figure}

The second reason is that the electronic state in a solid is specified by the wavenumber \(\mathbf{K}\). The potential energy of an object in real space is a function of position \(x, y, z\). On the other hand, the energy of an electron in a solid is a function of \(k_{x}, k_{y}, k_{z}\), which is the coordinates of reciprocal space. By investigating the relationship between such energy \(\mathbf{E}\) and wavenumber \(\mathbf{K}\), the electronic state in a solid can be understood.

In solid state physics, we mainly target crystals in which atoms are arranged periodically [\ref{fig:crystal-arragement}]. When the atoms are arranged periodically, the density of the electrons contained in them also becomes periodic, so it can be expressed as a wave.

\begin{figure}[htbp!]
    \centering
    \includegraphics[width=0.8\linewidth]{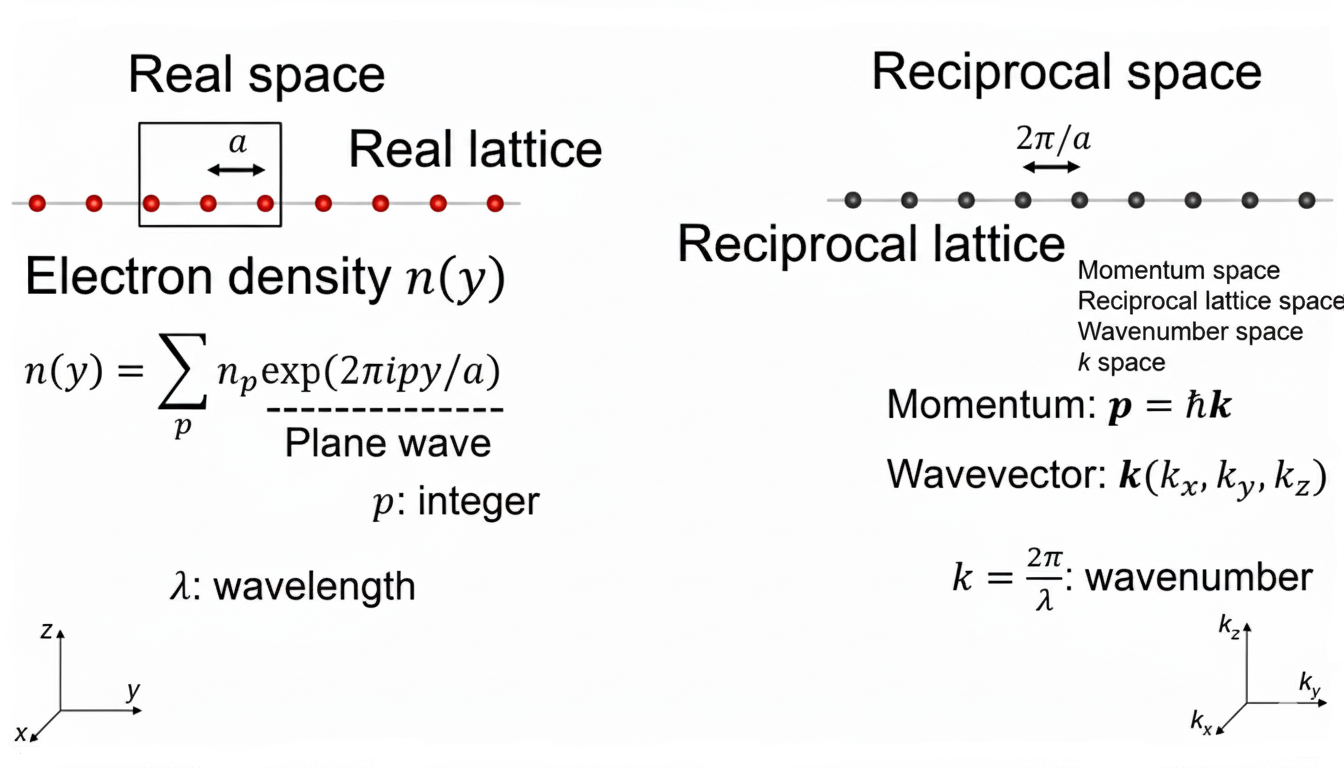}
    \caption{Crystal arragement}
    \label{fig:crystal-arragement}
\end{figure}

If the atoms are one-dimensionally arranged at intervals \(a\), the wave \(n(y)\) representing the electron density distribution can be represented by the sum of the waves with wavelength \(\lambda = a\). In the case of complex $3D$ structures and potentials, the structure can be expressed by superimposing waves with various wavelengths \(\lambda\). The lambda is the wavelength of any wave in three-dimensional real space. For wavelength \(\lambda\), \(2\pi/\lambda\) is defined as the wavenumber \(\mathbf{K}\). The wavenumber \(\mathbf{K}\) equals \(2\pi/\lambda\) represents the number of waves per unit length, and once the wavelength \(\lambda\) is determined, the wavenumber \(\mathbf{K}\) is also determined.

Since the atoms of a crystal are arranged three-dimensionally, the wavenumber \(\mathbf{K}\) also becomes a three-dimensional wavevector and has \(k_{x}, k_{y}, k_{z}\) as a component. The space whose coordinates are \(k_{x}, k_{y}, k_{z}\) is a reciprocal space. There are a reciprocal lattice and reciprocal space that reflects the periodicity of the real lattice corresponding to the crystal structure. The reciprocal lattice period is \(2\pi/\lambda\). 

It is called a reciprocal lattice or reciprocal space because the unit is the reciprocal of length \(a\). The momentum \(p\) is the wavenumber \(\mathbf{K}\) multiplied by Dirac constant. Therefore, the reciprocal space is also called the momentum space or the reciprocal lattice space, wavenumber space or \(\mathbf{K}\) space.

As described above, the structure and electronic state of a solid become periodic according to the periodic arrangement of atoms, so it is convenient to describe using waves.

\section*{Definition of Reciprocal Lattice}

Let \(a_{1}, a_{2}, a_{3}\) be the primitive vectors corresponding to the axes of the unit cell. Correspondingly, in the reciprocal space whose coordinates are \(k_{x}, k_{y}, k_{z}\). The primitive vectors of the reciprocal lattice are \(b_{1}, b_{2}, b_{3}\) [\ref{fig:primitive-vectors}]. At this time, there is such a relationship between \(a_{1}, a_{2}, a_{3}\) and \(b_{1}, b_{2}, b_{3}\). This is the definition of the primitive vector of the reciprocal lattice.

\begin{figure}[htbp!]
    \centering
    \includegraphics[width=0.8\linewidth]{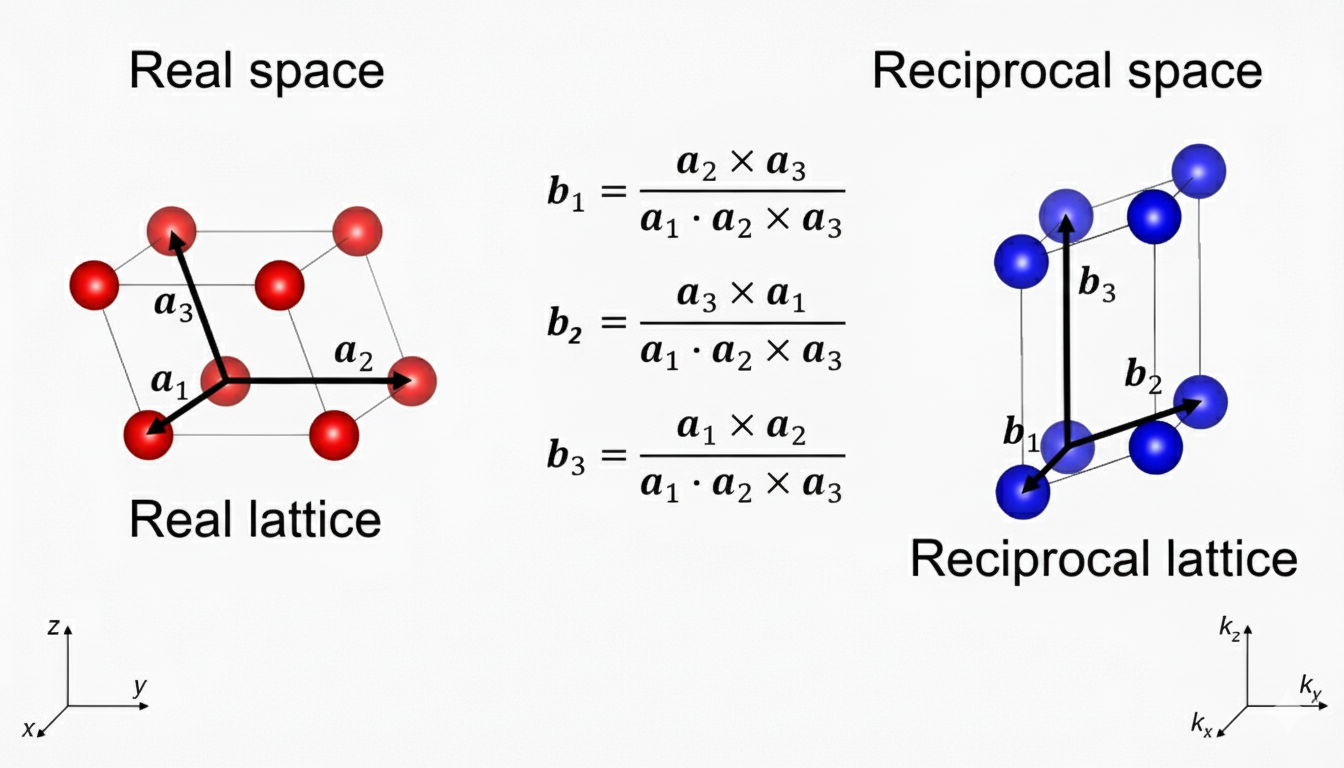}
    \caption{Correspondence of primitive vectors to reciprocal lattice}
    \label{fig:primitive-vectors}
\end{figure}

The denominator of these equations is the volume of the parallelepiped or unit cell created by \(a_{1}, a_{2}, a_{3}\). For example, the vector \(b_{1}\) in the reciprocal space is proportional to the cross product of \(a_{2}\) and \(a_{3}\), so its direction is perpendicular to the plane created by \(a_{2}\) and \(a_{3}\). Similarly, the vector \(b_{2}\) is perpendicular to the plane created by \(a_{3}\) and \(a_{1}\), and \(b_{3}\) is perpendicular to the plane created by \(a_{1}\) and \(a_{2}\). The magnitude of \(b_{1}\) is the reciprocal of the spacing of the planes made by \(a_{2}\) and \(a_{3}\). In this way, when the real lattice is defined, the reciprocal lattice is also defined.

For example, consider the case of a two-dimensional hexagonal honeycomb lattice [\ref{fig:honeycomb-to-reciprocal}]. When the primitive lattice vector is \(a_{1}\) and \(a_{2}\), the primitive vector of the reciprocal lattice \(b_{1}\) points in the direction perpendicular to the plane created by \(a_{2}\). After we determine the origin of the reciprocal lattice, we draw \(b_{1}\) diagonally downward to the right from the origin. The length of \(b_{1}\) is the reciprocal of the plane spacing, but first write it arbitrarily. The vector \(b_{2}\) has the direction perpendicular to the plane created by \(a_{1}\). The length of \(b_{2}\) is also the reciprocal of the plane spacing, but it is the same length as \(b_{1}\) and the hexagonal lattice. After drawing \(b_{1}\) and \(b_{2}\), draw \(-b_{1}\) and \(b_{2}\). Also draw \(b_{1}, b_{2}, -b_{1} + b_{2}\), etc.

\begin{figure}[htbp!]
    \centering
    \includegraphics[width=0.8\linewidth]{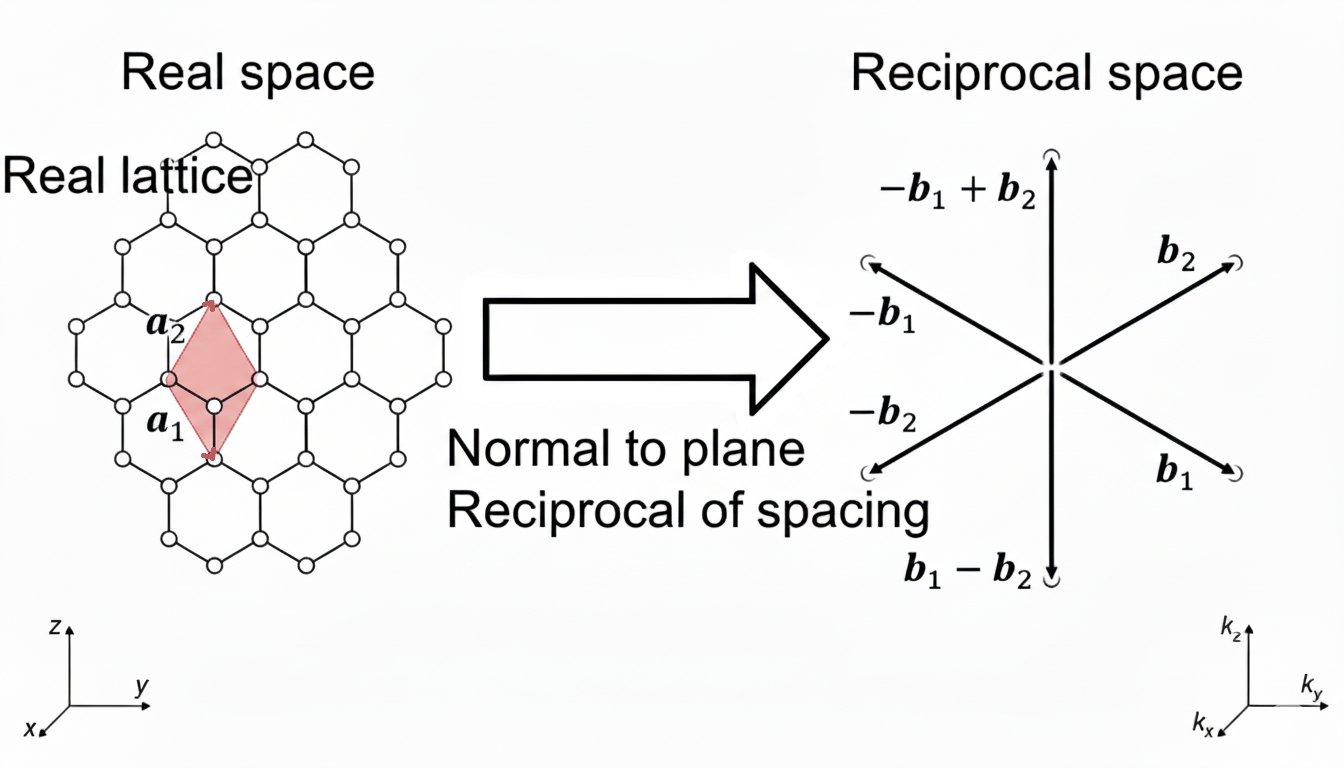}
    \caption{Two-dimensional hexagonal honeycomb lattice to reciprocal lattice}
    \label{fig:honeycomb-to-reciprocal}
\end{figure}

The lattice created in this way is the reciprocal lattice. If the lattice is a hexagonal lattice, the reciprocal lattice is also a hexagonal lattice. As there is a unit cell on the crystal structure, there is also a unit cell in the reciprocal lattice. This is called the Brillouin zone. The hexagon surronded by vertical bisectors of vectors \(b_{1}\) and \(b_{2}\) is called the first Brillouin zone. The highly symmetric points in the Brillouin zone are named [\ref{fig:brillouin-zone}]. For example, the origin is called the gamma (\(\Gamma\)) point, the hexagonal vertex is called the \(\mathbf{K}\) point, and the midpoint of the edge is called the endpoint.  In this way, when the real lattice is determined, which is an element of the crystal structure, the reciprocal lattice is also determined accordingly.  Conversely, once the reciprocal lattice is understood, the real lattice can be known.

\begin{figure}
    \centering
    \includegraphics[width=0.8\linewidth]{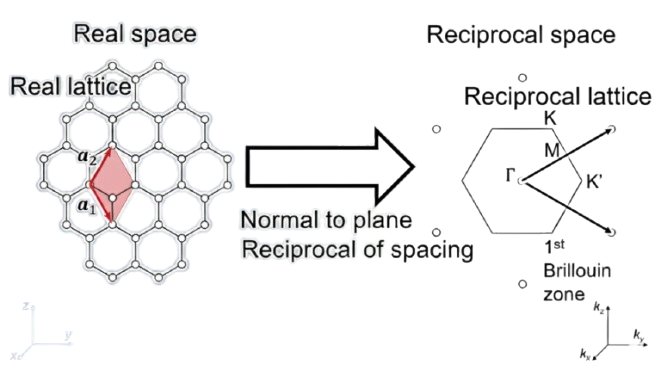}
    \caption{Brillouin zone}
    \label{fig:brillouin-zone}
\end{figure}

The reciprocal lattice can be indirectly observed in the x-ray diffraction experiment. However, in electron diffraction experiments, it is possible to directly observe the reciprocal lattice. An electron diffraction pattern can be obtained by using a transmission electron microscope. The electron diffraction pattern is an observation of the cross section of the reciprocal lattice.

The transformation between the real lattice and the reciprocal lattice is mathematically called the Fourier transformation. In the Fourier transformation equation \(f(x)\), which is a function of the coordinates \(x\) in the real space is transformed into \(f(k)\), which is a function of the coordinates \(k\) in the reciprocal space.
\section*{Reciprocal Lattice Used To Know The Electronic State of a Solid}

Many properties of solids are determined by their electronic state. The atoms in the solid are regularly arranged. Then the electrons in the crystal exist under periodic potential. Therefore, the electronic state depends on the wavenumber corresponding to the periodicity of the crystal. 

For example, the energy of an electron is a solid is a function of the wavenumber \(k\), which is the coordinates of the reciprocal space. The relation between energy and wavenumber in a solid is called the dispersion relation, and is often called the energy band structure. As an example,in the silicon band structure shown in the figure, the vertical axis is energy and the horizontal axis is wavenumber \(k\). The silicon is a semi-conductor because there is a region where energy does not exist for any wavenumber, that is a forbidden band. The energy width of the forbidden band is called the energy gap. 

Once we know the energy band structure, we can also understand the motion of electrons in solids.

According to Newton's equation of motion, when a force \(F\) is applied to an object with mass \(m\) in real space, the velocity \(v\) changes with time. The velocity \(v\) can be expressed as the time derivative of position \(x\). On the other hand, in solid state physics, when a force \(F\) is applied to an electron in a solid, the wavenumber \(k\) changes with time and the group velocity, which is the velocity of electrons, is expressed by the wavenumber derivative of energy. The wavenumber derivative of energy corresponds to the slope of the energy band. Here is the vicinity of the Fermi energy in the silicon band structure.

\chapter{Bloch's Theorem}
The Bloch's Theorem is fundamental in the study of periodic potentials, especially in the context of solid-state physics and the behavior of electrons in a crystal lattice. It states that the wave functions for electrons in a periodic potential can be written as a plane wave modulated by a periodic function with the same periodicity as the lattice.

To make more clear this concept, we star by defining the periodic potencial of the lattice. For example, we consider an electron in a crystal lattice where the potential \( V(\mathbf{r}) \) is periodic with the periodicity of the lattice. This means:

\begin{equation}
    V(\mathbf{r} + \mathbf{R}) = V(\mathbf{r}),
\end{equation}

\noindent for any lattice vector \(\mathbf{R}\).

The time-independent Schrödinger equation for an electron in this potential is:

\begin{equation}
    \left[ -\frac{\hbar^2}{2m} \nabla^2 + V(\mathbf{r}) \right] \psi(\mathbf{r}) = E \psi(\mathbf{r}).
\end{equation}

Introducing the translation operator \( \hat{T}_{\mathbf{R}} \), which translates a function by a lattice vector \(\mathbf{R}\):

\begin{equation}
    \hat{T}_{\mathbf{R}} \psi(\mathbf{r}) = \psi(\mathbf{r} + \mathbf{R}).
\end{equation}

Since the potential is periodic:

\begin{equation}
    \hat{T}_{\mathbf{R}} V(\mathbf{r}) \psi(\mathbf{r}) = V(\mathbf{r} + \mathbf{R}) \psi(\mathbf{r} + \mathbf{R}) = V(\mathbf{r}) \hat{T}_{\mathbf{R}} \psi(\mathbf{r}).
\end{equation}

The Hamiltonian \(\hat{H}\) commutes with the translation operator:

\begin{equation}
    \left[ \hat{H}, \hat{T}_{\mathbf{R}} \right] = 0.
\end{equation}

Since \(\hat{H}\) and \(\hat{T}_{\mathbf{R}}\) commute, they can share a common set of eigenfunctions. Let \(\psi(\mathbf{r})\) be an eigenfunction of \(\hat{T}_{\mathbf{R}}\):

\begin{equation}
    \hat{T}_{\mathbf{R}} \psi(\mathbf{r}) = \psi(\mathbf{r} + \mathbf{R}) = \lambda_{\mathbf{R}} \psi(\mathbf{r}).
\end{equation}

The eigenvalue \(\lambda_{\mathbf{R}}\) must have the form \( \lambda_{\mathbf{R}} = e^{i \mathbf{k} \cdot \mathbf{R}} \), where \(\mathbf{k}\) is the crystal momentum. This follows from the periodicity and the need for \(\lambda_{\mathbf{R}}\) to be consistent with translations by all lattice vectors.

The wavefunction can thus be written as:

\begin{equation}
    \psi(\mathbf{r} + \mathbf{R}) = e^{i \mathbf{k} \cdot \mathbf{R}} \psi(\mathbf{r}).
\end{equation}

This implies that the wavefunction can be expressed as:

\begin{equation}
    \psi(\mathbf{r}) = e^{i \mathbf{k} \cdot \mathbf{r}} u_{\mathbf{k}}(\mathbf{r}),
\end{equation}

\noindent where \( u_{\mathbf{k}}(\mathbf{r}) \) is a function with the same periodicity as the lattice:

\begin{equation}
    u_{\mathbf{k}}(\mathbf{r} + \mathbf{R}) = u_{\mathbf{k}}(\mathbf{r}).
\end{equation}

This theorem tells us that the solutions to the Schrödinger equation in a periodic potential are plane waves modulated by a periodic function. This form greatly simplifies the study of electron behavior in crystals and is foundational for understanding band theory and electronic properties of materials.

\chapter{Spacial Vectors Lattice Changed}
This appendix provides the definitions and Cartesian coordinate representations of the key spatial vectors used to describe the honeycomb lattice of graphene. The lattice constant $a \approx 1.42 \, \si{\angstrom}$ represents the nearest-neighbor carbon-carbon distance.

\section{Real-Space Lattice Vectors}
\label{app:real_space_vectors}

The honeycomb lattice of graphene can be described by a triangular Bravais lattice with a two-atom basis. The primitive lattice vectors $\vec{a}_1$ and $\vec{a}_2$ define the unit cell in real space (Eq. 3.1):
\begin{align}
\vec{a}_{1} &= \frac{a}{2}(3,\sqrt{3}) = \left( \frac{3a}{2}, \frac{a\sqrt{3}}{2} \right) \\
\vec{a}_{2} &= \frac{a}{2}(3,-\sqrt{3}) = \left( \frac{3a}{2}, -\frac{a\sqrt{3}}{2} \right)
\end{align}
Any lattice point in the Bravais lattice can be reached by integer linear combinations of these two vectors.

\section{Nearest-Neighbor Vectors}
\label{app:nn_vectors}

In the honeycomb structure, each carbon atom has three nearest neighbors. Assuming an atom of sublattice A is at the origin, the vectors pointing to its three nearest neighbors (belonging to sublattice B) are given by $\vec{\delta}_1$, $\vec{\delta}_2$, and $\vec{\delta}_3$ (Eq. 3.2):
\begin{align}
\vec{\delta}_{1} &= \frac{a}{2}(1,\sqrt{3}) = \left( \frac{a}{2}, \frac{a\sqrt{3}}{2} \right) \\
\vec{\delta}_{2} &= \frac{a}{2}(1,-\sqrt{3}) = \left( \frac{a}{2}, -\frac{a\sqrt{3}}{2} \right) \\
\vec{\delta}_{3} &= a(-1,0) = \left( -a, 0 \right)
\end{align}
These vectors are crucial for defining nearest-neighbor hopping terms in tight-binding models. Note that $|\vec{\delta}_1| = |\vec{\delta}_2| = |\vec{\delta}_3| = a$.

\section{Reciprocal Lattice Vectors}
\label{app:reciprocal_vectors}

The reciprocal lattice is fundamental for describing wave phenomena in the crystal, such as electron wave vectors $\vec{k}$. The primitive reciprocal lattice vectors $\vec{b}_1$ and $\vec{b}_2$ define the Brillouin zone and are related to the real-space vectors by $\vec{a}_i \cdot \vec{b}_j = 2\pi \delta_{ij}$. For the graphene lattice, they are given by (Eq. 3.3):
\begin{align}
\vec{b}_{1} &= \frac{2\pi}{3a}(1,\sqrt{3}) = \left( \frac{2\pi}{3a}, \frac{2\pi\sqrt{3}}{3a} \right) \\
\vec{b}_{2} &= \frac{2\pi}{3a}(1,-\sqrt{3}) = \left( \frac{2\pi}{3a}, -\frac{2\pi\sqrt{3}}{3a} \right)
\end{align}
These vectors define the basis for the reciprocal lattice, which is also a triangular lattice, rotated by $30^\circ$ relative to the real-space lattice. The first Brillouin zone constructed from these vectors is hexagonal.

\chapter{Relationship Between the Magnetic Field $\mathbf{B}$ and the Vector Potential $\mathbf{A}$}
\label{appendice:d}
In many physical systems, particularly in electromagnetism and fluid dynamics, it is common to represent a vector field $\mathbf{B}$ as the \textbf{curl} of another vector field $\mathbf{A}$, that is:

\begin{equation}
    \mathbf{B} = \nabla \times \mathbf{A}
\end{equation}

This equation encapsulates a fundamental relationship between the vector field $\mathbf{B}$ and the \textbf{vector potential} $\mathbf{A}$. In what follows, we break down and explain each term in this expression and provide an interpretation of the significance of $\mathbf{A}$ in physical and mathematical terms.

\section{The Curl Operator: $\nabla \times$}

The symbol $\nabla \times$ denotes the \textbf{curl} of a vector field. Mathematically, the curl of a vector field $\mathbf{A} = (A_x, A_y, A_z)$ is another vector field given by:

\begin{equation}
    \nabla \times \mathbf{A} = 
\begin{vmatrix}
\hat{\mathbf{i}} & \hat{\mathbf{j}} & \hat{\mathbf{k}} \\
\frac{\partial}{\partial x} & \frac{\partial}{\partial y} & \frac{\partial}{\partial z} \\
A_x & A_y & A_z
\end{vmatrix}
\end{equation}

The curl measures the \textbf{rotation} or \textbf{circulation} of the vector field $\mathbf{A}$ around a given point. Physically, it tells us how much and in what direction the field is ``curling'' around that point.

\section{The Magnetic Field $\mathbf{B}$}

In the context of classical electromagnetism, $\mathbf{B}$ represents the \textbf{magnetic field}, a vector field that describes the magnetic influence on moving electric charges, electric currents, and magnetic materials. According to Maxwell’s equations, specifically:

\begin{equation}
    \nabla \cdot \mathbf{B} = 0
\end{equation}

the magnetic field has zero divergence, implying that there are no magnetic monopoles---magnetic field lines always form closed loops.

This zero-divergence condition allows us to express $\mathbf{B}$ as the curl of another vector field $\mathbf{A}$, because the divergence of a curl is always zero:

\begin{equation}
  \nabla \cdot (\nabla \times \mathbf{A}) = 0  
\end{equation}

This mathematical identity ensures that the representation $\mathbf{B} = \nabla \times \mathbf{A}$ is always consistent with Maxwell's equations.

\section{The Vector Potential $\mathbf{A}$}

The vector field $\mathbf{A}$ is known as the \textbf{magnetic vector potential}. It is a mathematical construct from which the magnetic field $\mathbf{B}$ can be derived. While $\mathbf{A}$ is not directly observable, it plays a crucial role in the formulation of electromagnetic theory, especially in:

\begin{itemize}
    \item \textbf{Gauge theory}, where different choices of $\mathbf{A}$ (related through a gauge transformation) yield the same physical magnetic field $\mathbf{B}$.
    \item \textbf{Quantum mechanics}, particularly in the Aharonov--Bohm effect, where the vector potential affects the phase of the wave function even in regions where $\mathbf{B} = 0$.
    \item \textbf{Electromagnetic wave propagation}, where both $\mathbf{A}$ and the scalar potential $\phi$ are used to express the electric and magnetic fields.
\end{itemize}

\section{Physical Interpretation of the Relationship}

The equation $\mathbf{B} = \nabla \times \mathbf{A}$ tells us that the magnetic field is entirely determined by the local rotational behavior of the vector potential $\mathbf{A}$. This formulation is particularly useful for solving problems where the magnetic field must satisfy certain boundary conditions or when working in specific gauges (such as the Coulomb or Lorenz gauge).

Moreover, since the magnetic field is derived from the curl of $\mathbf{A}$, it is \textbf{invariant} under the transformation:

\begin{equation}
    \mathbf{A} \rightarrow \mathbf{A}' = \mathbf{A} + \nabla \Lambda
\end{equation}

for any scalar field $\Lambda$, due to the identity:

\begin{equation}
    \nabla \times (\nabla \Lambda) = \mathbf{0}
\end{equation}

This property reflects the \textbf{gauge freedom} of the vector potential and underlines its role in modern field theories.


\end{apendicesenv}





\phantompart
\printindex

\end{document}